\documentclass[%
 reprint,
 superscriptaddress,
 amsmath,amssymb,
]{revtex4-2}

\usepackage{braket}
\usepackage{graphicx}% Include figure files
\usepackage{dcolumn}% Align table columns on decimal point
\usepackage{bm}% bold math
\usepackage{hyperref}% add hypertext capabilities
\usepackage{amsmath}
\usepackage{graphicx}
\usepackage{tikz}
\usetikzlibrary{arrows.meta,positioning,calc}

\makeatletter
\renewcommand\p@subsection{\thesection.}
\makeatother

\makeatletter
\newcommand{\sublabel}[1]{%
  \label{#1}%
  \protected@write\@auxout{}{%
    \string\newlabel{#1@subonly}{{\arabic{subsection}}{\thepage}}%
  }%
}
\makeatother

\newcommand{\subnumref}[1]{\hyperref[#1]{\ref*{#1@subonly}}}

\begin{document}

\title{Crystal Orbital Guided Iteration to Atomic Orbitals: A Pathway
to Chemically Adaptive Atomic Orbitals from DFT}

\author{Emily Oliphant}
 \email{eoliphan@mit.edu}
\affiliation{
 Department of Materials Science and Engineering,
University of Michigan, Ann Arbor, Michigan 48109, United States
}
\affiliation{
 Schwarzman College of Computing, Massachusetts
Institute of Technology, Cambridge, MA 02139, United States
}%
\affiliation{
 Research Laboratory of Electronics, Massachusetts
Institute of Technology, Cambridge, MA 02139
}%
\author{Emmanouil Kioupakis}
\author{Wenhao Sun}
 \email{whsun@umich.edu}
\affiliation{
 Department of Materials Science and Engineering,
University of Michigan, Ann Arbor, Michigan 48109, United States
}

\begin{abstract}
Atomic orbitals underpin our understanding of electronic structure, unlocking intuitive descriptions of bonding, charge transfer, magnetism, and correlation effects. 
However, these descriptions are only reliable if the chosen atomic orbitals form a complete basis for the first-principles wavefunctions. 
Although Maximally localized Wannier functions (MLWFs) can form a complete orbital basis, the resulting Wannier orbitals smear onto neighboring atoms, obscuring their real-space atomic character. 
Here, we show that the deviation from atomic character in projected Wannier constructions arises from two intrinsic mathematical obstacles: uncontrolled orbital mixing and an approximate fixed-overlap constraint. 
To overcome these limitations, we introduce Crystal Orbital Guided Iteration To atomic-Orbitals (COGITO), a framework that iteratively adapts the underlying atomic orbitals such that the nonorthogonal Wannier representation achieves accurate tight-binding interpolation while maintaining atomic character. 
Our method thus offers a powerful tool for physics- or chemistry-based applications that rely on a faithful description of local electronic structure.
\end{abstract}

\maketitle

%\tableofcontents

\section*{Introduction}

The atomistic origin of electronic structure is of longstanding
ambiguity in condensed matter physics and quantum chemistry. While
atomic orbitals form an interpretable and compact basis for the
Kohn-Sham wavefunctions used in density functional theory (DFT), they
limit DFT from reaching the lowest energy state when electrons favor a
basis that is distorted from the atomic orbitals of the isolated atom.
In contrast, plane-waves paired with the projector augmented wave (PAW)
method form a complete basis for the Kohn-Sham wavefunctions,\cite{blochl_projector_1994}
which easily allows redistribution of the Kohn-Sham wavefunctions to
minimize energy, but this plane-wave basis obscures the spatial and
chemical character of the crystal orbitals.

To retain advantages from both atomic bases and plane-wave bases,
plane-wave wavefunctions can be projected onto atomic orbitals. These
projections have enabled local orbital-based electronic structure
post-processing methods such as tight-binding (TB)
interpolation\cite{sanchez-portal_projection_1995,gresch_automated_2018}, crystal orbital Hamilton populations
(COHP)\cite{dronskowski_crystal_1993}, atomic charge decomposition\cite{mulliken_electronic_1955}, spin-orbit
coupling\cite{steiner_calculation_2016}, and many-body corrections including DFT+U
\cite{anisimov_first-principles_1997,mosey_ab_2007,agapito_reformulation_2015,shih_screened_2012}, magnetic exchange \cite{liechtenstein_local_1987,antropov_exchange_1997,szilva_interatomic_2013,solovyev_exchange_2021}, and dynamical mean-field
theory \cite{georges_dynamical_1996,kotliar_electronic_2006,haule_exact_2015}. However, the choice of projected basis and
possible augmentation of the projections (often orthonormalization)
strongly affects the fidelity and interpretability of such applications.
Although a wide variety of strategies have been developed to construct
local orbitals and their projections, each guided by its own design
principles\cite{sanchez-portal_projection_1995,marzari_maximally-localized_1997,ozaki_variationally_2003,qian_quasiatomic_2008,hoyvik_orbital_2012,sakuma_symmetry-adapted_2013,knizia_intrinsic_2013,wang_selectively_2014,damle_compressed_2015,agapito_accurate_2016,maintz_lobster_2016,heselmann_local_2016,jonsson_theory_2017,damle_disentanglement_2018,damle_variational_2019,folkestad_implementation_2022,ozaki_closest_2024,schreder_propagated_2024}, no existing approach simultaneously achieves
completeness, locality, and chemical interpretability.

Here we propose four criteria to guide the assessment of local orbital
bases in achieving a transferable and chemically predictive description
of electronic structure. \textbf{(1)} The basis should be
\textbf{chemically interpretable}, carrying the form of atomic orbitals
without distortion from the angular function of spherical harmonics.
\textbf{(2)} The basis should be \textbf{adaptable and unique},
capturing shifts in orbital radial function under different charge
states and crystal environments while being independent of how the
projection is initialized. \textbf{(3)} The \textbf{basis should span
the Kohn-Sham wavefunctions} (and vice versa), meaning the unmodified
projections should satisfy completeness of Kohn-Sham bands and the
projected orbitals---where completeness is quantified by the charge
spillage\cite{sanchez-portal_projection_1995} and orbital mixing (a term we introduce later).
\textbf{(4)} The basis must enable \textbf{high-quality tight-binding}
interpolation, such that the orbital projections give rise to
tight-binding models with \textless10 meV valence band error compared to
DFT.

Any minimal set of predefined atomic orbitals can be chemically
interpretable (Criterion 1). However, since these bases do not adapt, they do not sufficiently span the
Kohn--Sham states and have poor tight-binding interpolation. A variety
of techniques strive to produce local orbitals that better span the
Kohn-Sham wavefunctions and produce good tight-binding interpolations
(Criteria 3 and 4), but come with limitations in adaptability,
uniqueness, and chemical interpretability (Criteria 1 and 2). Early
approaches searched for an optimized orbital basis by restricting the
orbital to a functional form with tunable parameters.\cite{chadi_localized-orbital_1977,eschrig_optimized_1988} For
example, Sanchez-Portal \emph{et al.} defines the optimized basis as the
PAW pseudo-orbital multiplied by the scale factor that achieves the
lowest charge spillage.\cite{sanchez-portal_projection_1995} Unfortunately, the ambiguity in
selecting a functional form, paired with rigid constraints on the
orbital shape, limits the reliability and adaptability of the basis.
Furthermore, while band interpolation may be improved with these optimized
bases, band errors can often be \textgreater{} 1 eV for a minimal
basis.\cite{sanchez-portal_analysis_1996}

To achieve more accurate band interpolations, the projection matrices can
be symmetrically orthonormalized to ensure completeness (either of the
Kohn-Sham bands and/or the projected orbitals). Then, variational
strategies like Maximally Localized Wannier Functions
(MLWFs)\cite{marzari_maximally-localized_1997,marzari_maximally_2012,qiao_projectability_2023} can be used to further augment the projections by
optimizing the locality or another desired feature of resulting Wannier
functions.\cite{wannier_structure_1937} However, modifying the projection matrices decouples
them from the orbital basis used in projection, reducing the atomic
interpretability of the modified projection matrices, Hamiltonian
matrices, and Wannier functions.

In particular, enforcing orbital orthogonality forces the resulting
orbitals to mix with atomic orbitals on neighboring atoms, making
orthogonal orbitals non-transferable between systems.\cite{chan_highly_2007,artacho_nonparametrized_1991,koepernik_symmetry-conserving_2023} As
demonstrated by Chan \emph{et. al.} in Figures 9--11\cite{chan_highly_2007}, the
Hamiltonian elements of an orthogonal basis no longer reflect simple
atomic overlap energy but instead encode nonlocal features. This arises
from system-dependent oscillating tails, where Wannier orbitals 
improperly smear onto neighboring atoms to maintain orthogonality.
Consequently, COHP and atomic-based analyses derived from an
orthogonalized basis\cite{maintz_lobster_2016,kundu_population_2021,taylor_pengwann_2025} should be interpreted with caution.

Quasi-atomic orbitals (QO)\cite{qian_quasiatomic_2008,chan_highly_2007} and nonorthogonal generalized
Wannier functions (NGWFs)\cite{skylaris_nonorthogonal_2002,oregan_subspace_2011} are closer to a strictly atomic
basis by working within a nonorthogonal framework. However, the orbitals
still deviate from atomic interpretability by exhibiting tails around
neighboring atoms\cite{qian_quasiatomic_2008}, although these nonlocal tails are smaller
than in orthogonal functions. In \textbf{Sec.~\ref{sec:distort}}, we analyze the
origin of these distortions and show that projected nonorthogonal
Wannier functions lose atomic interpretability through two distinct
mechanisms---orbital mixing and an implicit fixed-overlap
constraint---leaving the resulting basis strongly dependent on
initialization.

To remedy these limitations, we use nonorthogonal Wannier functions to
guide the construction of a strictly atomic orbital basis
(\textbf{Criterion 1}), which in turn refines a Wannier representation
that preserves atomic character. We introduce our scheme as
\emph{Crystal Orbital Guided Iteration To atomic-Orbitals} (COGITO) in
\textbf{Sec.~\ref{sec:create}}. The central idea in COGITO is to perform iterative,
chemically guided modifications to the Wannier representation that break
the fixed-overlap constraint and suppress orbital mixing. The full
COGITO process---including the Bloch orbital update, coefficient
refinement, and atomic orbital fitting---is iterated such that the
atomic orbital basis and its overlap matrix converge to a chemically
faithful representation of the KS electronic structure.

In \textbf{Sec.~\ref{sec:compare}}, we evaluate how the COGITO basis adapts to
chemical environments on a test set of 200 semiconductors and metals
\cite{vitale_automated_2020}, finding a tenfold reduction in sensitivity to basis
initialization \textbf{(Criterion 2)}. \textbf{Sec.~\ref{sec:ham}} outlines the
construction of a nonorthogonal tight-binding model from COGITO.
Lastly, \textbf{Sec.~\ref{sec:results}} demonstrates that the COGITO basis accurately spans
the Kohn-Sham wavefunctions \textbf{(Criterion 3)} and yields
high-quality tight-binding interpolations of band structure
\textbf{(Criterion 4)}. 

Readers interested in a comparison of COGITO with MLWF plus projectability detangled Wannier functions (PDWF) are directed to
the end of \textbf{Sec.~\ref{sec:results}}.
The full open-source COGITO package, covering atomic basis construction through bonding analysis, is available via our webpage, Ref. \cite{oliphant_cogito_2026}.

\section{The origin of distortions in projected Wannier orbitals}
\label{sec:distort}

Projected Wannier functions have been used extensively, yet their
distortion from the projected orbital basis is not a common discussion
point in literature. Understanding and controlling this distortion is
essential in guiding the creation of an optimal basis. We find that the
projected nonorthogonal Wannier functions distort from the original
projected basis in two distinct ways. First, when the Kohn-Sham (KS)
wavefunctions do not form a complete set for the projected orbitals, the
orbital character of the Wannier functions becomes a mix of multiple
projected orbitals. We name this undesirable distortion `orbital
mixing'. Second, when the projected basis does not form a complete set
for the KS wavefunctions, the resulting projected Wannier functions are
modified from the initial atomic basis. While this adaptation is crucial
to span the KS wavefunctions, we identify that this update occurs under
an approximate fixed-overlap constraint that unnecessarily delocalizes the projected
Wannier functions onto neighboring atoms. Together, these distortions
drive nonorthogonal Wannier functions away from chemical
interpretability (Criterion 1), adaptability and uniqueness (Criterion
2).

\subsubsection*{\textbf{General and projected Wannier constructions}}

A general Wannier function is constructed as the Fourier transform of
Bloch periodic states. In the simplest case, the periodic states are the
KS wavefunctions \(\psi_{n}^{\mathbf{k}}\), where \(n\), \(\mathbf{k}\), and
\(\mathbf{R}\) are the KS band, \textbf{k}-point, and lattice translation
vector, respectively.

\begin{equation}
\ket{\psi_{n}^{\mathbf{R}}}
= \sum_{\mathbf{k}} e^{- i \mathbf{k}\cdot \mathbf{R}} \ket{\psi_{n}^{\mathbf{k}}}.
\tag{1}
\end{equation}

The gauge freedom in the band Wannier functions
\(\psi_{n}^{\mathbf{R}}\) (i.e. that \(\psi_{n}^{\mathbf{k}}\) has an
arbitrary complex phase at each \textbf{k}-point) can be fixed by
projecting atomic Bloch orbitals \(\Phi_{\alpha}^{\mathbf{k}}
= \sum_{\mathbf{R}} e^{i\mathbf{k}\cdot(\mathbf{R}+\boldsymbol{\tau}_{\alpha})}\,
\phi_{\alpha}^{\mathbf{R}+\boldsymbol{\tau}_{\alpha}}\), where \(\phi_{\alpha}^{\mathbf{R}+\boldsymbol{\tau}_{\alpha}}\) is an atomic orbital centered at \(\mathbf{R}+\boldsymbol{\tau}_{\alpha}\). This
creates the projected Wannier functions \(\phi_{\alpha}'^{\ \mathbf{R}+\boldsymbol{\tau}_{\alpha}}\),
which \emph{should be} similar in character to the local atomic
orbitals.

\begin{equation}
\ket{\phi_{\alpha}'^{\ \mathbf{R}+\boldsymbol{\tau}_{\alpha}}}
= \sum_{\mathbf{k}} e^{- i \mathbf{k}\cdot (\mathbf{R}+\boldsymbol{\tau}_{\alpha})}
\sum_{n} \ket{\psi_{n}^{\mathbf{k}}}\braket{\psi_{n}^{\mathbf{k}}|\Phi_{\alpha}^{\mathbf{k}}}.
\tag{2}
\end{equation}

Whether the complex phase of the orbital center in the primitive cell, \(\boldsymbol{\tau}_{\alpha}\), is included explicitly in constructing the Bloch orbitals and Wannier orbitals is a matter of convention.\cite{yusufaly2013tight} While most do not include the phase, we do so for subtle reasons discussed near the end of \textbf{Sec.~\ref{sec:ham_ham}}. 

For orthogonal Wannier functions, the transformation matrix
\(\braket{\psi_{n}^{\mathbf{k}}|\Phi_{\alpha}^{\mathbf{k}}}\)
must be unitary, but this is not necessary for a generalized
nonorthogonal construction. The indices \(\alpha\) and \(\beta\)
indicate atomic states (local or Bloch) while \(n\) and \(m\) indicate
KS states. The number of atomic states can be less than the number of KS
states, i.e. the transformation matrix may be rectangular.

\subsubsection*{\textbf{Decomposition of KS states into atomic and residual
components}}

The KS wavefunctions can be decomposed into a component captured by the
atomic Bloch orbitals, and a residual component that lies outside the
atomic basis.

\begin{equation}
\ket{\psi_{n}^{\mathbf{k}}}
= \sum_{\alpha\beta}\ket{\Phi_{\alpha}^{\mathbf{k}}}\,{S^{\mathbf{k}}_{\beta\alpha}}^{-1}\,\braket{\Phi_{\beta}^{\mathbf{k}}|\psi_{n}^{\mathbf{k}}}
+ \ket{\Delta\psi_{n}^{\mathbf{k}}}.
\tag{3}
\end{equation}

where
\(S_{\beta\alpha}^{\mathbf{k}} = \braket{\Phi_{\beta}^{\mathbf{k}}|\Phi_{\alpha}^{\mathbf{k}}}\)
is the atomic Bloch orbital overlap. For an orthogonal basis,
\(S_{\alpha\beta}\) is the identity matrix but can be any complex Hermitian
matrix for our nonorthogonal basis.

The first part of \textbf{Eqn. 3} is equivalent to representing \(\psi_{n}^{\mathbf{k}}\)
as a linear combination of \(\Phi_{\alpha}^{\mathbf{k}}\), with the coefficients
\(c_{\alpha n}^{\mathbf{k}}\) (\textbf{Eqn. 4}) describing the amount of orbital
\(\alpha\) in band \(n\).

\begin{equation}
c_{\alpha n}^{\mathbf{k}}
= \sum_{\beta} {S^{\mathbf{k}}_{\beta\alpha}}^{-1}\,\braket{\Phi_{\beta}^{\mathbf{k}}|\psi_{n}^{\mathbf{k}}}.
\tag{4}
\end{equation}

The second term of \textbf{Eqn. 3}, \(\mathrm{\Delta}\psi_{n}^{\mathbf{k}}\), represents
the residual wavefunction---the component of the KS state that cannot be
expressed in the atomic basis and is therefore orthogonal to the atomic
Bloch orbitals
(\(\braket{\Phi_{\beta}^{\mathbf{k}}|\mathrm{\Delta}\psi_{n}^{\mathbf{k}}} = 0\)).
If the atomic basis is inadequate, this residual can be substantial even
for low-energy KS states. By contrast, an ideal atomic basis should
achieve no residual for the occupied KS states. This residual is later quantified as band spillage and is also the complement to projectability.

For another perspective, let us define the subspace $\mathcal{K}$ to span the KS states and the subspace $\mathcal{A}$ to span the atomic Bloch orbitals. There are an infinite possible number of KS states, so let us define two limited KS subspaces: $\mathcal{K}_{occ}$ to span the occupied KS states, and $\mathcal{K}_{inc}$ to span all included KS states. In practice, which KS states are included can be determined by the number of the KS state, the energy of the KS state, or the band spillage of the KS state. 

In this terminology, Eqn. 3 shows the projection of elements of $\mathcal{K}_{inc}$ onto elements of $\mathcal{A}$ and $\mathcal{A}^{\perp}$, where projection onto $\mathcal{A}^{\perp}$ becomes the residual wavefunction.
Then, our criteria for an ideal atomic basis (that elements of $\mathcal{K}_{occ}$ have no residual wavefunction) indicates $\mathcal{K}_{occ} \subseteq \mathcal{A}$. 
In practice, this is only approximate.

\subsubsection*{\textbf{Orbital mixing and fixed-overlap constraint in projected
Wannier functions}}

Now we can determine how projected Wannier functions differ from the
initial atomic orbitals by examining their periodic counterparts,
\(\Phi_{\alpha}'^{\ \mathbf{k}}\) and \(\Phi_{\alpha}^{\mathbf{k}}\).

\begin{equation}
\ket{\Phi_{\alpha}'^{\ \mathbf{k}}}
= \sum_{n = 1}^{N}\ket{\psi_{n}^{\mathbf{k}}}\,\braket{\psi_{n}^{\mathbf{k}}|\Phi_{\alpha}^{\mathbf{k}}}.
\tag{5}
\end{equation}

The projected Bloch orbital \(\Phi_{\alpha}'^{\ \mathbf{k}}\) is identical
to the atomic Bloch orbitals \(\Phi_{\alpha}^{\mathbf{k}}\) when the
included KS states (of subspace $\mathcal{K}_{inc}$) span all of the atomic Bloch orbitals (of subspace $\mathcal{A}$), 
such that $\mathcal{A} \subseteq \mathcal{K}_{inc}$. 
%Under this condition, all states in $\mathcal{A}$ can be espressed as a linear combination
%of the states in subspace $\mathcal{K}_{inc}$ without loss of information.
When $\mathcal{A} \nsubseteq \mathcal{K}_{inc}$, the portion of 
\(\Phi_{\alpha}^{\mathbf{k}}\) that is orthogonal to the subspace $\mathcal{K}_{inc}$ is excluded from \(\Phi_{\alpha}'^{\ \mathbf{k}}\), such that $\mathcal{A}' \subseteq \mathcal{K}_{inc}$.

This deviation is quantified by substituting \textbf{Eqn. 3} in for
\(\ket{\psi_{n}^{\mathbf{k}}}\) in \textbf{Eqn. 5}. We
introduce \(\xi\) as an additional orbital index and write sums with
implicit matrix multiplications. For brevity, we drop the \textbf{k}-index.

\begin{equation}
\ket{\Phi_{\alpha}'}
= \ket{\Phi_{\beta}}\,{S_{\xi\beta}}^{-1}\,\braket{\Phi_{\xi}|\psi_{n}}\braket{\psi_{n}|\Phi_{\alpha}}
+ \ket{\Delta\psi_{n}}\,\braket{\psi_{n}|\Phi_{\alpha}}.
\tag{6}
\end{equation}

\textbf{Eqn. 6} shows the projected Bloch orbital \(\Phi_{\alpha}'\) is distorted
from the original atomic Bloch orbitals in two ways, which we simplify
by defining the \emph{orbital mixing matrix}
\(M_{\beta\alpha}^{\mathbf{k}}\) and the updated orbital
\(\ket{\mathrm{\Delta}\Phi_{\alpha}}\).
%An ideal atomic basis, with subspace $\mathcal{A} \subseteq \mathcal{K}_{inc}$, would already have an orbital mixing matrix that is the identity matrix and a $\Delta\psi_{n}$ that is zero.

\begin{equation}
\ket{\Phi_{\alpha}'}
= \ket{\Phi_{\beta}}\,M_{\beta\alpha} + \ket{\Delta\Phi_{\alpha}}.
\tag{7}
\end{equation}

Where:

\begin{equation}
M_{\beta\alpha}
\equiv {S_{\xi\beta}}^{-1}\,\braket{\Phi_{\xi}|\psi_{n}}\braket{\psi_{n}|\Phi_{\alpha}}
= c_{\beta n}\,c_{\xi n}^{\dagger}\,S_{\xi\alpha},
\tag{8}
\end{equation}
\begin{equation}
\ket{\Delta\Phi_{\alpha}}
\equiv \ket{\Delta\psi_{n}}\,\braket{\psi_{n}|\Phi_{\alpha}}
= \ket{\Delta\psi_{n}}\,c_{\xi n}^{\dagger}\,S_{\xi\alpha}.
\tag{9}
\end{equation}

The first term mixes the original atomic Bloch orbitals when the included KS wavefunctions do not form a complete basis. % for the atomic Bloch orbitals. 
If only a few KS bands are included (i.e. only valence bands), then the mixing matrix will specify how the resulting Wannier functions form hybrid orbitals.\cite{Foster_NHO_1980,Dunnington_periodicNHO_2012}
But when many KS states are included, we postulate that it is possible to achieve $\mathcal{A}^{i+1} \subseteq \mathcal{K}_{inc}$ by changing only the shape of each orbital, where $\mathcal{A}^{i+1}$ is the subspace that spans these updated orbitals.
This is supported by the high variance in off-diagonal components of \(M_{\beta\alpha}\) with changes in the shape of initial basis, 
such as implementing a cutoff radius or shrinking the basis, see \textbf{Table I}. 
%The mixing matrix is also sensitive to excluding high-energy KS bands with atomic character. 
In this scenario, explicitly mixing the orbitals should be avoided and all orbital shape update should be supplied by the second term (although $M_{\beta\alpha}=\mathbb{I}$ generally indicates $\Delta\Phi_{\alpha}=0$, since both occur when $\mathcal{A} \subseteq \mathcal{K}_{inc}$). 

\begin{table}
\caption{\label{tab:max-mixing}Maximum off-diagonal orbital mixing from projecting onto different orbital bases.}
\begin{ruledtabular}
\begin{tabular}{cc}
Ti$_2$Ag & $\max\!\left(\bm{M}_{\bm{\beta}\neq\bm{\alpha}}\right)$ \\
\hline
COGITO basis & 0.0075 \\
PAW pseudo + exp fit & 0.0387 \\
20\% smaller PAW & 0.0805 \\
50\% smaller PAW & 0.1589 \\
20\% larger PAW & 0.0664 \\
50\% larger PAW & 0.4495 \\
PAW with cutoff at 1.5~\AA & 0.1841 \\
PAW with cutoff at RDEPT (1.952 \& 2.072) & 0.0686 \\
\end{tabular}
\end{ruledtabular}
\end{table}

However, the second term, \textbf{Eqn. 9}, is where an approximate fixed-overlap constraint emerges.
The \(\ket{{\Delta}\Phi_{\alpha}}\) term
necessarily updates the projected Bloch orbitals such that they are accurately spanned by
the KS wavefunctions but may deviate from the perfect atomic character
of the initial basis. 
%Later (Sec. II.2), we also ensure selected KS states are accurately spanned by \(\Phi_{\alpha}'\) by modifying the projection matrices.
%In essence, this ensures $\mathcal{A}' \subseteq \mathcal{K}_{inc}$ and can ensure $\mathcal{K}_{sel} \subseteq \mathcal{A}'$. But $\mathcal{A}'$ may deviate from $\mathcal{A}$ in a way that is uncessary/unatomic. 
%One other important fact is that this subspaces are perfectly spanning for one k-point but once combined may deviate slightly from perfect spanning. 
But because the residual wavefunctions are
strictly orthogonal to the initial atomic Bloch basis
(\(\braket{\Phi_{\beta}|\Delta\psi_{n}} = 0\)),
any update constructed from them must also be orthogonal to all initial
orbitals (i.e.
\(\braket{\Phi_{\beta}|\Delta\Phi_{\alpha}} = 0\)).
As a result, these changes in the projected Bloch orbitals are inherently limited in changing the orbital overlap,
enforcing an approximate fixed-overlap constraint. The constraint is approximate since 
the overlap can be perturbed by the orbital mixing and \(\braket{\Delta\Phi_{\beta}|\Delta\Phi_{\alpha}}\).
While not as severe as a full orthogonality, this approximate fixed-overlap
constraint leads to nodal tails and deviation from perfectly atomic
character by mixing with neighboring atoms.

\begin{figure}[t!]
\includegraphics[width=3.3in]{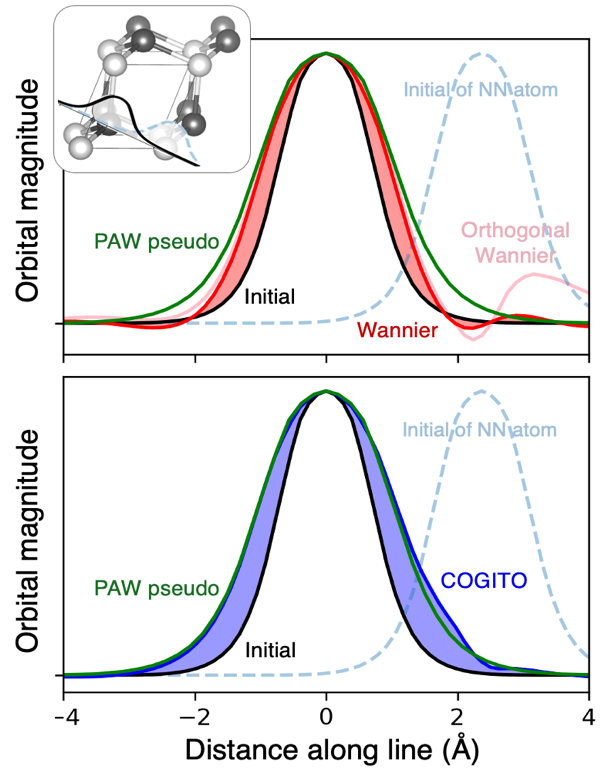}
\caption{Demonstrates the ability of Wannier (top) and COGITO
(bottom) to adapt back to the PAW pseudo-orbital when initialized with
30\% shrunken orbitals. Plots the silicon s-like orbital on a 1D slice
between neighboring atoms, shown in the top left schematic.}
\label{fig:distort}
\end{figure}

In \textbf{Fig.~\ref{fig:distort}}, we demonstrate the effects of the approximate fixed-overlap constraint by
plotting the \emph{s}-like projected Wannier function created from 30\%
shrunken PAW pseudo-orbitals. The orthogonal \emph{s}-like Wannier
function (pink) has a large oscillating tail by the neighboring atom.
Since these distortions arise from the requirement of zero overlap with
the neighboring Wannier functions, the oscillating tail is similar for
any size of projected basis. While nonorthogonal Wannier functions (red)
have a reduced oscillating tail, they are still incapable of flexing
back to the original PAW shape due to the constraint that updates to the
Wannier orbital (shaded red region) must be orthogonal to the
surrounding projected orbitals (dashed gray). This approximate fixed-overlap
constraint also makes the shape of the nonorthogonal tail heavily
dependent on the size of the starting basis, as it defines the majority of the
final orbital overlap.

The requirement of \(\mathrm{\Delta}\Phi_{\alpha}\) to have limited impact on
the overlap is also observed from the direct calculation of the projected
Bloch function overlap using \textbf{Eqn. 5} and substituting from \textbf{Eqn. 8} to
reach the right hand side.

\begin{align*}
\braket{\Phi_{\xi}'|\Phi_{\alpha}'}&=\braket{\Phi_{\xi}|\psi_{m}} \braket{\psi_{m}|\psi_{n}} \braket{\psi_{n}|\Phi_{\alpha}} \\
&=\braket{\Phi_{\xi}|\psi_{m}}\delta_{mn}\braket{\psi_{n}|\Phi_{\alpha}} = S_{\xi\beta}M_{\beta\alpha}.
\end{align*}

In their construction of a nonorthogonal quasi-atomic basis, T.L. Chan
\emph{et. al.} observed the consequences of the approximate fixed-overlap constraint
remarking ``\emph{Since the QUAMBOs are deformed according to different
bonding environments, it is expected that the overlap integrals can vary
for different crystal structures. However, from Fig. 7, \textbf{the
overlap integrals corresponding to different structures fall onto the
same curve very nicely}}.'' The overlaps across multiple structures
neatly falling on one curve is an artifact of this constraint, where the
overlaps curve is strictly defined by the orbital basis chosen for
projection. Small deviations from the initial overlap are a consequence
of \(M_{\beta\alpha} \neq \mathbb{I}\).

%\clearpage
\section{Creating COGITO}
\label{sec:create}

Our central aim is to construct Wannier functions that are maximally
atomic to extract chemical insight from the Kohn--Sham (KS) electronic
structure. 
A natural starting point for obtaining a highly atomic
Wannier representation is by projecting atomic orbitals (\textbf{Eqn. 2}), which
constructs the Wannier functions that most closely resemble the initial
atomic orbitals within the Hilbert space of the included KS states. As
shown in \textbf{Sec.~\ref{sec:distort}}, however, these projected Wannier functions can still
deviate from perfect atomic character due to orbital mixing and an
implicit fixed-overlap constraint. Even so, they typically represent a
step towards an atomic description that more faithfully spans the
included KS subspace. As such, we suggest an iterative strategy that
alternates between constructing projected Wannier functions and using
them to fit a new atomic basis, allowing the Wannier and atomic
representations to co-evolve. This strategy illustrates the essence of
COGITO but faces two key challenges: the resulting Wannier functions are
not guaranteed to span the desired KS states, and the naive loop is
highly computationally inefficient.

We address these challenges through our Crystal Orbital Guided Iteration
To atomic-Orbitals (COGITO) scheme, which efficiently finds a highly
atomic Wannier representation that spans the desired KS subspace. COGITO
breaks the approximate fixed-overlap constraint by iteratively updating our orbital
basis to span the KS wavefunctions while performing chemically-guided
modifications to the orbitals and enforcing that they remain strictly
atomic. Our procedure is broken into three key steps:

\begin{quote}
\textbf{\ref{sec:create_atomic}} Restrict the updated Bloch orbitals to have no orbital
mixing and the correct complex phase of reciprocal-space coefficients.

\textbf{\ref{sec:create_opt}} Modify the orbital coefficients to span the KS wavefunctions
as desired.

\textbf{\ref{sec:create_fit}} Extract numerical local orbitals from the numerical Bloch
orbitals at \textbf{k} = 0 and fit to analytical atomic orbitals in a
flexible multi-Gaussian form.
\end{quote}

\textbf{Steps 1} and \textbf{3} are geared towards finding the best
atomic orbital for the projected Wannier functions in a computationally
trackable manner. \textbf{Step 1} restricts the formulation of the
projected Bloch orbital to remove orbital mixing, nudging the Wannier
representation towards atomic character. Later, \textbf{Step 3}
explicitly fits local atomic orbitals to radial Gaussian functions while
preserving an exponential-like decay in the orbital tail. 
By introducing a Hirshfeld-like partitioning scheme to extract local orbitals using only the $\Gamma$-point KS states, we avoid the high computational cost of constructing local Wannier orbitals from a dense \textbf{k}-point grid.

\begin{figure}[t!]
\resizebox{\columnwidth}{!}{%
\begin{tikzpicture}[>=Stealth,
    line width=1pt,
    boxblue/.style={
        draw=cyan!70!blue!95!black!100,
        rounded corners=4pt,
        align=center,
        inner sep=6pt,
        fill=black!6
    },
    boxorange/.style={
        draw=orange!95!red!80!black!100,
        rounded corners=4pt,
        align=center,
        inner sep=6pt,
        fill=black!6
    },
    boxyellow/.style={
        draw=orange!30!yellow!90!black!100,
        rounded corners=4pt,
        align=center,
        inner sep=6pt,
        fill=black!6
    },
    boxgreen/.style={
        draw=green!75!blue!70!black!100,
        rounded corners=4pt,
        align=center,
        inner sep=4pt,
        fill=black!3
    },
    blackarrow/.style={-{Stealth[length=6pt,width=8pt]},draw=black,line width=2pt},
    redarrow/.style={-{Stealth[length=6pt,width=8pt]},draw=cyan!70!blue!95!black!100,line width=2pt},
    greenarrow/.style={-{Stealth[length=6pt,width=7pt]},draw=green!75!blue!70!black!100,line width=1.5pt}
]

% Top box
\node[boxyellow] (pw) at (0,0)
    {Plane-wave DFT\\ calculation};

% Second box
\node[boxblue] (init) at (0,-1.40)
    {Initialize orbitals\\ from PAW data};

% Third box
\node[boxblue] (proj1) at (0,-2.80)
    {Project orbitals on\\ KS wavefunctions};

% Large middle box
\node[boxblue, line width=2pt] (modify) at (-0.0,-4.50)
    {\bfseries Modify projected Bloch orbital\\[2pt]
     \ref{sec:create_atomic} Update to be more atomic\\
     \ref{sec:create_opt} Update to span select KS bands};

% Right middle orange box
\node[boxorange, line width=2pt] (extract) at (5.0,-4.50)
    {\bfseries \ref{sec:create_fit} Extract atomic\\ \bfseries gaussian orbitals};

% Lower right blue box
\node[boxblue] (proj2) at (4.55,-6.20)
    {\ref{sec:ham_proj} Project orbitals \\
     \ref{sec:ham_sym} Expand to full \textbf{k}-grid};

% Lower left blue box
\node[boxblue] (build) at (-0.0,-6.20)
    {\ref{sec:ham_ham} Build \textbf{k/R}-dependent\\ orbital Hamiltonians};

% Bottom orange box
\node[boxorange] (output) at (0.0,-7.7)
    {Output overlap and\\ hopping parameters};
% Bottom orange box

\node[boxgreen] (analyze1) at (4.4,-1.)
    {\ref{sec:compare} Analyze orbital changes \\  and check sensitivity};
    
% Bottom orange box
\node[boxgreen] (analyze2) at (3.8,-7.70)
    {\ref{sec:results} Verify quality};

% Black arrows
\draw[blackarrow] (pw.south) -- (init.north);
\draw[blackarrow] (init.south) -- (proj1.north);
\draw[blackarrow] ($(proj1.south)-(0.60,0)$) -- ($(modify.north)-(0.60,0)$);
\draw[blackarrow] (modify.east) -- (extract.west);
\draw[blackarrow] ($(extract.south)-(0.60,0)$) -- ($(proj2.north)-(0.15,0)$);
\draw[blackarrow] (proj2.west) -- (build.east);
\draw[blackarrow] (build.south) -- (output.north);

% Red feedback arrows
\draw[redarrow] 
    ($(modify.north)+(0.60,0)$) -- ($(proj1.south)+(0.60,0)$);

\draw[redarrow]
    ($(extract.north)-(0.60,0)$) .. controls +(0,1.2) and +(1.0,-0.) .. (proj1.east);

\draw[greenarrow] ($(extract.north)-(0.0,0)$) -- ($(analyze1.south)+(0.60,0)$);
\draw[greenarrow] (output.east) -- (analyze2.west);

\end{tikzpicture}
}
\caption{Illustrates the overall workflow of COGITO and highlights the sections
in this paper that discuss various steps.}
\label{fig:diagram}
\end{figure}

%\begin{figure}[t!]
%\includegraphics[width=3.3in]{media/image1-2.png}
%\caption{Illustrates the overall workflow of COGITO and highlights the %sections
%in this paper that discuss various steps.}
%\end{figure}

\textbf{Step 2} ensures the Wannier representation spans the desired KS
states. In standard Wannier constructions, this is achieved by
user-defined `frozen' and `inclusion' windows, which specify the bands
to be symmetrically orthogonalized to enforce completeness and the bands
included in construction, respectively. The choice of frozen and
inclusion windows often controls the quality of the Wannier
representation and involves difficult tradeoffs, particularly when band
accuracy (favors freezing desired bands with a narrow inclusion window)
and atomic character (favors including all high-energy bands with atomic
projection) must both be retained. To address this, in \textbf{Step 2} we propose
a combined Gram-Schmidt and Lowdin orthogonalization that enforces
completeness for the valence bands while hierarchically decoupling bands
by energy and projectability. The lowest-energy valence bands are Lowdin
orthogonalized first and remain invariant during subsequent Gram-Schmidt
orthogonalization of higher-energy valence and conduction bands. As a
result, the quality of low-energy bands is preserved even as additional
bands are frozen to improve band accuracy or included to reduce orbital
mixing, thereby reducing sensitivity to window selection.

The COGITO procedure alternates between refining atomic-like Bloch
orbitals (\textbf{Steps 1} and \textbf{2}) and fitting corresponding strictly atomic
orbitals (\textbf{Step 3}). The iteration over \textbf{Steps 1} and \textbf{2} yields numerically
defined atomic-like Bloch orbitals that span the valence KS states while
preserving atomic character. Convergence of the self-consistent loop is
monitored through the band spillage, which directly reflects convergence
of the projected orbital coefficients and their corresponding Bloch
orbitals. This loop is iterated until the band spillage converges to
within four decimal places, approaching the behavior outlined in \textbf{Sec.~\ref{sec:create_opt}} (negligible spill for valence bands and small spill for select
low-lying conduction bands). The full iteration over \textbf{Steps 1--3} then
fits these Bloch orbitals to analytical atomic orbitals. Because this
stage involves constrained least-squares fitting rather than a strict
variational minimum, convergence is assessed empirically: repeating the
full cycle four times is sufficient to achieve self-consistency
in orbital radii and overlap matrices to within 1--5\%. Overall, this
procedure is found to be robust in practice (see \textbf{Sec.~\ref{sec:compare}}).

Many Wannier schemes seek an optimal solution through variational
minimization of a chosen functional, generally written in terms of
projection or overlap matrices for computational efficiency. However,
``maximal atomicity'' does not correspond to a well-defined variational
target.
Here, we define ``maximally atomic'' as the Wannier orbitals that are maximally close to a separable description of radial and angular parts (i.e. \textbf{Eqn. 21}). 
Other works encourage atomic orbital character by enforcing orbital center or orbital symmetry\cite{wang_selectively_2014,sakuma_symmetry-adapted_2013}, but these definitions still permit extended orbital tails around neighboring atoms. In our definition, the optimal, perfectly separable atomic orbitals are themselves adaptable (through the radial functions) and are therefore not expressible as a fixed reference function for a variational target. Moreover, formulations
written purely in terms of projected matrices necessarily exclude the
portion of KS Hilbert space that lies outside the initial atomic
basis---the very component that must be incorporated for atomic orbitals
to evolve meaningfully across different chemical environments. Rather
than optimizing within a fixed representation defined by the initial
basis, COGITO iteratively updates both the Wannier representation and
the atomic orbitals used to make it.

As an illustrative example, the bottom of \textbf{Fig.~\ref{fig:distort}} shows
wavefunctions constructed from COGITO compared to projected Wannier
functions. Unlike the \emph{s}-like nonorthogonal Wannier function
(red), where the approximate fixed-overlap constraint causes an oscillating tail at
the neighboring orbital (dashed gray), COGITO (blue) bypasses the
fixed-overlap constraint and flexes back to the original PAW shape. This
highlights COGITO's ability to create an adaptable (\textbf{Criterion
2}) highly atomic (\textbf{Criterion 1}) Wannier representation for the
KS wavefunctions.

\subsection{Enforce atomic character in Bloch orbitals}
\label{sec:create_atomic}
\sublabel{subsec:create_atomic}

In COGITO, we iteratively update the orbitals with a component from the
residual KS wavefunctions. This approach aims to reduce orbital mixing
by setting \(M_{\alpha\beta} = \mathbb{I}\) in the projected Wannier
orbital (\textbf{Eqn. 7}) resulting in \textbf{Eqns. 10} and \textbf{11}. Mathematically, this is
akin to solving the linear equation of \textbf{Eqn. 5} in a self-consistent
linear iteration scheme, which becomes nonlinear from the additional
operations on \(\mathrm{\Delta}\Phi_{\alpha}^{i}\) and
\(c_{\xi n}^{i}\). We initialize our procedure with the PAW
pseudo-orbitals of the valence shell.\footnote{We are using the PBE\_52 pseudo-potentials, which define the radial
pseudo-orbitals up to $\sim$2-4 \AA\ for valence states. Despite this extended range, we still fit the tail to exponential
decay using the position and slope of the pseudo-orbital
at the end of its definition. This is especially relevant for excited state
\emph{p} orbitals, where the pseudo-orbital sharply drops to zero at the PAW cutoff.
The results in Sections \ref{sec:create} and \ref{sec:compare} use the PAW
pseudo-orbital with our fit exponential tail to better convey changes in the radial distribution. Whereas, results in Section \ref{sec:results} use the
PAW pseudo-orbitals without the exponential tail for better reproducibility and
accurate reflection of the PAW pseudo-orbital values.}

\begin{equation}
\ket{\Phi_{\alpha}^{i+1}} = \ket{\Phi_{\alpha}^{i}} + \ket{\Delta\Phi_{\alpha}^{i}},
\tag{10}
\end{equation}
\begin{equation}
\ket{\Delta\Phi_{\alpha}^{i}} = \sum_{n}\ket{\Delta\psi_{n}^{i}}\,{c_{\xi n}^{i}}^{\dagger}\,S_{\xi\alpha}^{i}.
\tag{11}
\end{equation}

Furthermore, behavior of the additional orbital,
\(\ket{\mathrm{\Delta}\Phi_{\alpha}^{i}}\), can be
restricted to ensure the orbital character of
\(\ket{\Phi_{\alpha}^{i + 1}}\). A convenient property
of Fourier transforms is that even functions have real Fourier
components while odd functions have imaginary Fourier components. Since
\emph{s} and \emph{d} atomic orbitals are even functions about the atom
center, we enforce the correct orbital character by requiring their
Fourier components to be real. Similarly, \emph{p} atomic orbitals are
odd functions, requiring imaginary Fourier components. \textbf{Equations 12} and
\textbf{13} show how we update the additional orbital in the plane-wave basis
\(\mathbf{G}\) and how the plane-wave coefficients are restricted to the
correct phase.

\begin{equation}
\ket{\Delta\Phi_{\alpha}} = \sum_{\mathbf{G}} c_{\mathbf{G}\alpha}\,\ket{e^{i\mathbf{G}\cdot \mathbf{r}}},
\tag{12}
\end{equation}
\begin{equation}
c_{\mathbf{G}\alpha} \equiv (-i)^{l}e^{- i \mathbf{G}\cdot \bm{\tau}_{\alpha}}\,
\mathrm{Re}\left[(-i)^{-l}e^{i\mathbf{G}\cdot \bm{\tau}_{\alpha}}c_{\mathbf{G}\alpha}\right].
\tag{13}
\end{equation}

Where \(l\) is the quantum number for angular momentum and the
\(e^{- i \mathbf{G}\cdot \bm{\tau}_{\alpha}}\) terms account for the phase shift from
the orbitals position in the primitive cell \(\bm{\tau}_{\alpha}\). By
ensuring the right phase of our atomic-like Bloch orbitals in Fourier
space, we prevent mixing of the orbital character with other orbitals on
both the same atom and neighboring atoms.

Running this section to self-consistent convergence guarantees only that
the updated atomic-like Bloch orbitals satisfy
\(M_{\alpha\beta}^{i + 1} = \mathbb{I}\), i.e. no additional orbital
mixing occurs since the included KS bands now span the updated subspace $\mathcal{A}^{i+1}$ (defined by the \(\Phi_{\alpha}^{i + 1}\) basis). In other terminology, $\mathcal{A}^{i+1} \subseteq \mathcal{K}_{inc}$. When the number of KS bands included
in \textbf{Eqn. 11} equals the number of orbitals,
\(M_{\alpha\beta}^{i + 1} = \mathbb{I}\) also implies that the Bloch
orbitals span the included KS bands. However, when more KS bands are
included (as is generally required to capture all relevant atomic character), the
updated Bloch orbitals are not guaranteed to span any of included KS
bands. Therefore, the next step is to refine the coefficient matrix such
that the KS bands are spanned in the desired way.

\subsection{Optimize coefficients}
\label{sec:create_opt}
\sublabel{subsec:create_opt}

Next, to enforce how the atomic-like Bloch orbitals span the KS
wavefunctions, we modify the projected orbital coefficient matrix used
in \textbf{Eqn. 11}. The band overlap matrix \(B_{nm}\) quantifies how accurately
each Kohn-Sham state is represented by the atomic orbitals:

\begin{equation}
B_{nm}
= \sum_{\alpha\beta}\braket{\psi_{n}|\Phi_{\alpha}}\,S^{-1}_{\alpha\beta}\,\braket{\Phi_{\beta}|\psi_{m}}
= c_{\alpha n}^{\dagger}\,S_{\alpha\beta}\,c_{\beta m}.
\tag{14}
\end{equation}

The diagonal of the band overlap matrix determines the widely used
metric of "band spillage" introduced by Sanchez-Portal et. al.\cite{sanchez-portal_projection_1995},
where \emph{P}\textsubscript{n} measures how much of each KS band
\emph{n} is lost when projected onto the atomic basis:

\begin{equation}
P_{n} = \mathrm{diag}\left(\mathbb{I}-B_{nm}\right).
\tag{15}
\end{equation}

The off-diagonal part of \(B_{nm}\) indicates a mixing of the KS bands
when downfolding to our minimal orbital basis and indicates that bands
with overlap will be incorrectly reproduced in the latter tight-binding model.

Before modifying the coefficient matrix, let us take a moment to consider what \(B_{nm}\) would be in different scenarios: (1) perfectly detangled bands, (2) fully entangled bands, and (3) entangled bands with lowest bands being perfectly described by basis. For detangled bands (1), \(B_{nm}\) should be the identity matrix. In subspace terminology, this corresponds to $\mathcal{K}_{iden} \subseteq \mathcal{A}$ where $\mathcal{K}_{iden}$ spans the KS states that form the identity matrix. In practice, the projected orbital set is not perfect, and a small deviation of \(B_{nm}\) from identity is commonly resolved by orthonormalizing the coefficient matrix.

For fully entangled bands (2), \(B_{nm}\) will not have any
restrictions, in fact, the matrix around high-energy bands is often far
from identity. While this seems undesirable, the variation from identity
correctly captures how the plane-wave solution downfolds onto the
minimal valence shell basis. Still an identity structure is often sought
after by mixing the KS wavefunctions to create a new \(B_{ij}\) that is
identity\cite{chan_highly_2007}, performing orthonormalization\cite{maintz_analytic_2013} to force
\(B_{nm} = \mathbb{I}\) for the M lowest (or highest projected) subset
of bands, or excluding any bands where \(B_{nm}\) varies too much from
identity\cite{agapito_effective_2013}. Alternatively, the band overlap can remain
unrestricted (\(B_{nm} \neq \mathbb{I}\)) by mapping onto atomic-like
Bloch orbitals that are correctly orthonormalizing under the KS
transformation\cite{marzari_maximally-localized_1997}. When generalized to a nonorthogonal basis, this
orbital orthogonality under KS transformation equates to the criterion
identified above that the orbital mixing matrix \(M_{\alpha\beta}\)
should be identity.

\begin{figure}[tb!]
\includegraphics[width=3.4in]{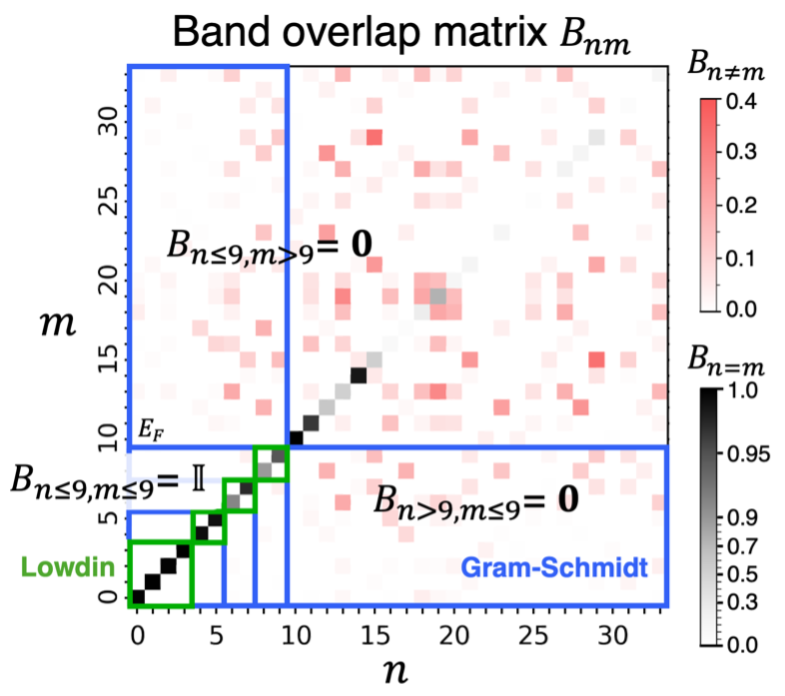}
\caption{Heat map of the band overlap matrix created from
Si\textsubscript{2}Ni KS wavefunctions projected on PAW
pseudo-orbitals. The low-energy section of \(B_{nm}\) is close to
identity, the mid-energy section is heavily mixed, and the high-energy
section approaches zero. The two-scale black colorbar for the diagonal
highlights small deviations from 1, while the red colorbar for
off-diagonal elements highlights deviations from 0. The partial-identity
structure to be enforced is shown mathematically. The green or blue
highlights represent sections to be Lowdin or Gram-Schmidt
orthogonalized.}
\label{fig:Bnm_matrix}
\end{figure}

Although referencing \(M_{\alpha\beta}\) to be identity instead of
\(B_{nm}\) can be very useful for tight-binding construction, this will
still produce interpolated bands that stray from the KS bands wherever \(B_{nm}\)
is not identity. To optimize the fit of our valence states, we consider
our third, most-physical, scenario (3): entangled bands with the
lowest-energy bands being perfectly described by the atomic orbital
basis. In this scenario, \(B_{nm}\) will be identity for a smaller
subset of bands but will be unrestricted for bands at higher energies.
\textbf{Fig.~\ref{fig:Bnm_matrix}} plots a heat map of \(B_{nm}\) from PAW
pseudo-orbitals in Si\textsubscript{2}Ni to visualize the matrix across
low-energy to high-energy regions. By enforcing a partial identity
construction, we will ensure the COGITO basis properly describes the
low-energy bands (below the Fermi energy) while allowing high-energy
bands to remix from the downfolding to atomic states.

To set the partial identity form in \(B_{nm}\), we start by selecting
the valence bands (bands 0-9 in \textbf{Fig.~\ref{fig:Bnm_matrix}}) as the low energy
region required to be identity. This can be achieved by Lowdin symmetric
orthogonalization. However, an overlap between a low-spill, low-energy
band and a mid-spill, mid-energy band would equally shift both bands
under symmetric orthogonalization, when ideally just the mid-spill band
would be adjusted to maintain orthogonality. While Gram-Schmidt
orthogonalization could be used, it distorts the symmetry of degenerate
bands.

To avoid both pitfalls, we devise a combined Lowdin + Gram-Schmidt
orthogonalization approach. It begins by grouping bands into multiple
sets based on the difference metric \(d_{n}\) defined below in \textbf{Eqn. 16},
grouping bands that are close in spillage or energy together, where $E_{scale}$ is selected as 1 eV. 

\begin{equation}
d_{n} = (P_{n+1}-P_{n})\,\tanh(\frac{E_{n+1}-E_{n}}{E_{scale}}).
\tag{16}
\end{equation}

COGITO goes through each band and adds the band to the current set if
\(d_{n}\) is less than the average \(d_{n}\) (for \(n\) in valence
bands), otherwise, it creates a new set. The selected sets are seen as
the green boxes in \textbf{Fig.~\ref{fig:Bnm_matrix}}. Then Lowdin orthogonalization is
performed on the lowest set (\(L\)) and all sets above (\(H\)) are
Gram-Schmidt orthogonalization to the resulting bands, as in \textbf{Eqns. 17}
and \textbf{18}. \(B_{nm}\) is updated after each step such
that \(B_{LL}\mathbf{=}1\) in \textbf{Eqn. 18}.

\begin{align}
\check{c}_{\alpha L}
&= (B^{-1/2})_{L'L}\,c_{\alpha L'},
\tag{17}\\
\check{c}_{\alpha H}
&= c_{\alpha H} - \frac{B_{HL}}{B_{LL}}\,\check{c}_{\alpha L}.
\tag{18}
\end{align}

This continues for all sets that are included in the identity region for
\(B_{nm}\). Finally, all bands outside the identity region are
Gram-Schmidt orthogonalized to the lower sets, see the large blue boxes
in \textbf{Fig.~\ref{fig:Bnm_matrix}}. This guarantees that the high bands are linearly
independent such that the low-energy bands will remain unaffected by any
remixed of the high-energy bands. With our approach, we preserve the
quality of low energy states while maintaining key symmetries in the
reconstructed KS states. For example, degenerate KS states will be grouped together and thus remain degenerate under symmetric Lowdin orthonormalization. For comparison of \(B_{nm}\) from the COGITO basis and after our
partial-orthonormalization procedure, see \textbf{Fig.~\ref{fig:band_error}} in \textbf{Sec.~\ref{sec:results}}.

Our hybrid scheme helps create an atomic basis that accurately captures
the essential chemical bonding characteristics of valence bands while
avoiding contributions from higher energy orbitals. In \textbf{Sec.~\ref{sec:ham_ham}}, we discuss how our orthogonalization scheme is tuned to create
tight-binding Hamiltonians for accurate interpolation of valence bands
and low energy conduction bands.

\subsection{Fit atomic orbitals from Bloch orbitals}
\label{sec:create_fit}
\sublabel{subsec:create_fit}

Next, we fit analytical atomic orbitals, \(\phi_{\alpha}\), to our basis
of atomic-like Bloch orbitals at \(\mathbf{k} = 0\), found
from the last section. These are related by \textbf{Eqn. 19}, which creates a
Bloch orbital from the sum over atomic orbitals in different translated
primitive cells times a phase factor. Shown later in \textbf{Eqn. 28},
constructing Bloch orbitals in a plane-wave basis does not even require
an explicit sum over translated cells.

\begin{equation}
\Phi_{\alpha}^{\mathbf{k}}(\mathbf{r})
= \sum_{\mathbf{R}} e^{i\mathbf{k}\cdot(\mathbf{R}+\boldsymbol{\tau}_{\alpha})}\,
\phi_{\alpha}\!\left(\mathbf{r}-\mathbf{R}-\boldsymbol{\tau}_{\alpha}\right).
\tag{19}
\end{equation}

While it is straightforward to build an atomic Bloch orbital from a local atomic orbital, it
can be challenging to decompose the local orbital that created a Bloch
orbital. Projected Wannier orbitals (\textbf{Eqn. 2}) achieve this by taking a Fourier transform of
the atomic Bloch orbitals (transformed from KS states) but are computationally expensive. As a
new alternative, we implement a simple algorithm to extract the atomic
orbital from a Bloch orbital. We introduce the ratio term
\(\chi_{\alpha}(r)\) in \textbf{Eqn. 20} that when multiplied by the Bloch
orbital at \(k = 0\), returns the atomic orbital. The ratio is
initialized using the PAW pseudo atomic orbitals with an exponential
decay and self-consistently iterated using the new fitted orbitals 0-4
times depending on the size of the unit cell. This is mathematically
similar to Hirshfeld partitioning of electron density\cite{hirshfeld_bonded-atom_1977}, but here
is used on the electron wavefunction.

\begin{equation}
\chi_{\alpha}(\mathbf{r})
= \frac{\phi_{\alpha}\!\left(\min_{\mathbf{R}}(\mathbf{r}-\mathbf{R}-\boldsymbol{\tau}_{\alpha})\right)}%
{\sum_{\mathbf{R}}\phi_{\alpha}\!\left(\mathbf{r}-\mathbf{R}-\boldsymbol{\tau}_{\alpha}\right)}.
\tag{20}
\end{equation}

\begin{figure}[htb!]
\includegraphics[width=3.4in]{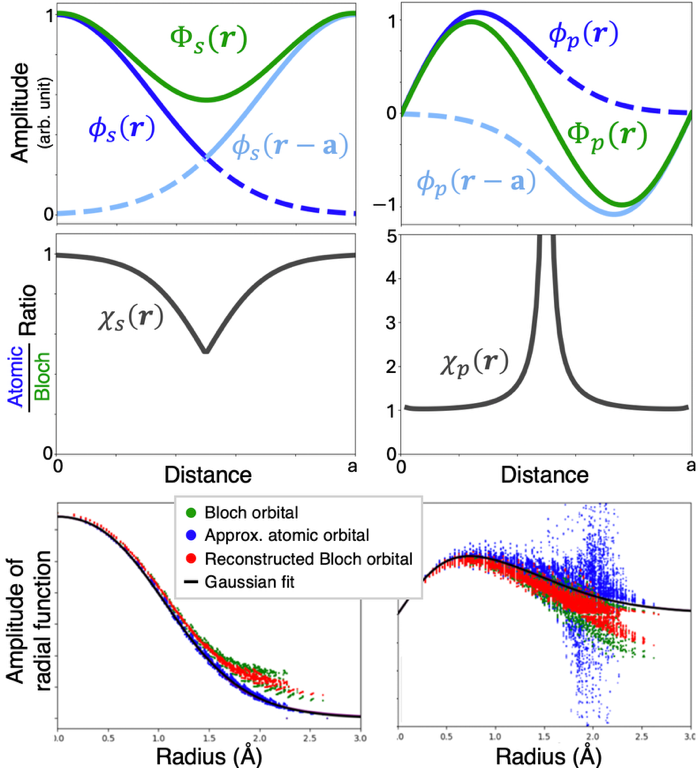}
\caption{Demonstration of atomic orbitals being distilled
from their atomic Bloch orbitals. The left column shows results for an s orbital
while the right shows a p orbital. Top row is a simple schematic for a 1D
primitive cell with the lines being colors dark blue, light blue, and green to
indicate the atomic orbital at R=0, the atomic orbital at R=1,
and the sum of them (the atomic Bloch orbital). The second row shows the ratio
defined in \textbf{Eqn. 20}, being the solid line for the atomic orbitals,
divided by the Bloch orbital. The final row shows these values for
silicon, where the green, blue, and red dots show the radial part for
the Bloch orbitals, approximate atomic orbital from the ratio, and the
reconstructed Bloch orbital after using the fitting radial function,
which is the black line.}
\label{fig:atomic_fit}
\end{figure}

Here,
\(\min_{R}\left( \mathbf{r} - \mathbf{R} - \bm{\tau}_{\alpha} \right)\)
indicates that each point throughout the periodic primitive cell has the
coordinates with respect to the closest periodic atom (by
\textbf{R}) to that point. Schematics for an \emph{s} orbital (left) and
\emph{p} orbital (right) are shown in \textbf{Fig.~\ref{fig:atomic_fit}}. The top row
plots the Bloch and atomic parts, while the middle row plots the ratio. 
Lastly, the bottom row plots the COGITO data for silicon in the diamond structure. 
Because an \emph{s} orbital does not have angular nodes, it only interferes
constructively with periodic images of itself in the \textbf{k} = 0 Bloch orbital, thus its ratio is a
well-behaved function. However, angular nodes in \emph{p} or
\emph{d} orbitals result in destructive interference between periodic images that cause the ratio to diverge whenever the Bloch orbital
is zero while the atomic orbital is not. Thus, points on the real-space
grid with a large or small ratio value are removed or given less weight
to improve the fit. %\footnote{}

Once a guess for the atomic orbital is found, the radial part is
separated by dividing out the real spherical harmonics for that orbital,
\textbf{Eqn. 21}. Then the radial part is fit to a sum of gaussians as
in \textbf{Eqn. 22}.

\begin{align}
\phi_{\alpha}(\mathbf{r})
&= R_{\alpha}^{l}(r)\,Y_{m}^{l}(\varphi,\theta),
\tag{21}\\
R_{\alpha}^{l}(r)
&= r^{l}\sum_{i} a_{\alpha}^{i}\,e^{-b_{\alpha}^{i}r^{2}}.
\tag{22}
\end{align}

Gaussians are selected over exponential functions to better reproduce
the behavior of PAW pseudo-wavefunctions at atom centers, which do not
possess the cusps at \(r = 0\) that exponential functions have. However,
to preserve the exponential decay of traditional atomic orbitals, the
numerical radial part outside a cutoff radius is first fit to
exponential functions before fitting the combined gaussian and
exponential radial function to gaussians. Fitting parameters for the
gaussians are constrained to prevent nodes while allowing a good fit.\footnote{Allowing a negative gaussian in \textbf{Eqn. 22} substantially improves fits for the COGITO orbitals because it allows for a more controlled decay (usually slower) from maximum while not overestimating the end tail. 
However, allowing a negative gaussian often leads to a long-range node in the radial function (often around 4 \AA). Even if the node is barely perceivable, it will completely throw off the desired decay of overlap and hopping terms. 
We identified that the following simple constraints to the coefficients in \textbf{Eqn. 22} mathematically prohibit a node. 
We define one positive gaussian to have the largest exponential factor $b^1$ (slowest decay) with associated $a^1>0$. Then, we define one possibly negative gaussian to have $b^{neg}<b^1$ and $a^{neg}+a^1>0$. 
Combined, these constraints ensure that a node does not appear in the radial function. This constraint is removed for semi-core states which have a node in the PAW pseudo-orbital.}
The bottom row of \textbf{Fig.~\ref{fig:atomic_fit}} shows the
standard output from COGITO to visualize the radial part of the orbitals
in silicon.

The key to COGITO lies in its self-consistent iteration: each cycle
constrains the orbitals to remain strictly atomic while adapting to
capture the DFT wavefunctions. This dual requirement captures the
essence of a chemically interpretable and accurate local basis, building
COGITO to satisfy our four criteria from the introduction. The
constraints applied in our iterative procedure suppresses unphysical
orbital mixing, enforces the correct orbital symmetry in Fourier-space,
refines the local orbital coefficients toward their optimal atomic form,
and finally promotes strictly atomic Bloch orbitals. This process
naturally breaks the approximate fixed-overlap constraint, allowing each orbital to
flex in shape and overlap while still preserving its atomic identity.
Combined, COGITO converts the static projection of DFT into a dynamic
and chemically interpretable atomic framework that reconstructs the
Kohn--Sham wavefunctions with both precision and purpose.

\section{How does COGITO compare to PAW pseudo-orbitals?}
\label{sec:compare}

To analyze the adaptability and independence \textbf{(Criterion 2)} of
COGITO, we run our workflow on a set of 200 nonmagnetic materials
previously benchmarked by Vitale \emph{et. al.}\cite{vitale_automated_2020} The set
includes 64 insulators/semiconductors and 136 metals. Our workflow to
run these uses the default pymatgen\cite{jain_high-throughput_2011} input parameters for a
static Vienna Ab initio Simulation Package (VASP)\cite{hafner_abinitio_2008} calculation,
but with a higher number of bands (12 bands per atom).

\begin{figure*}[hbt!]
\centering
\includegraphics[width=7in]{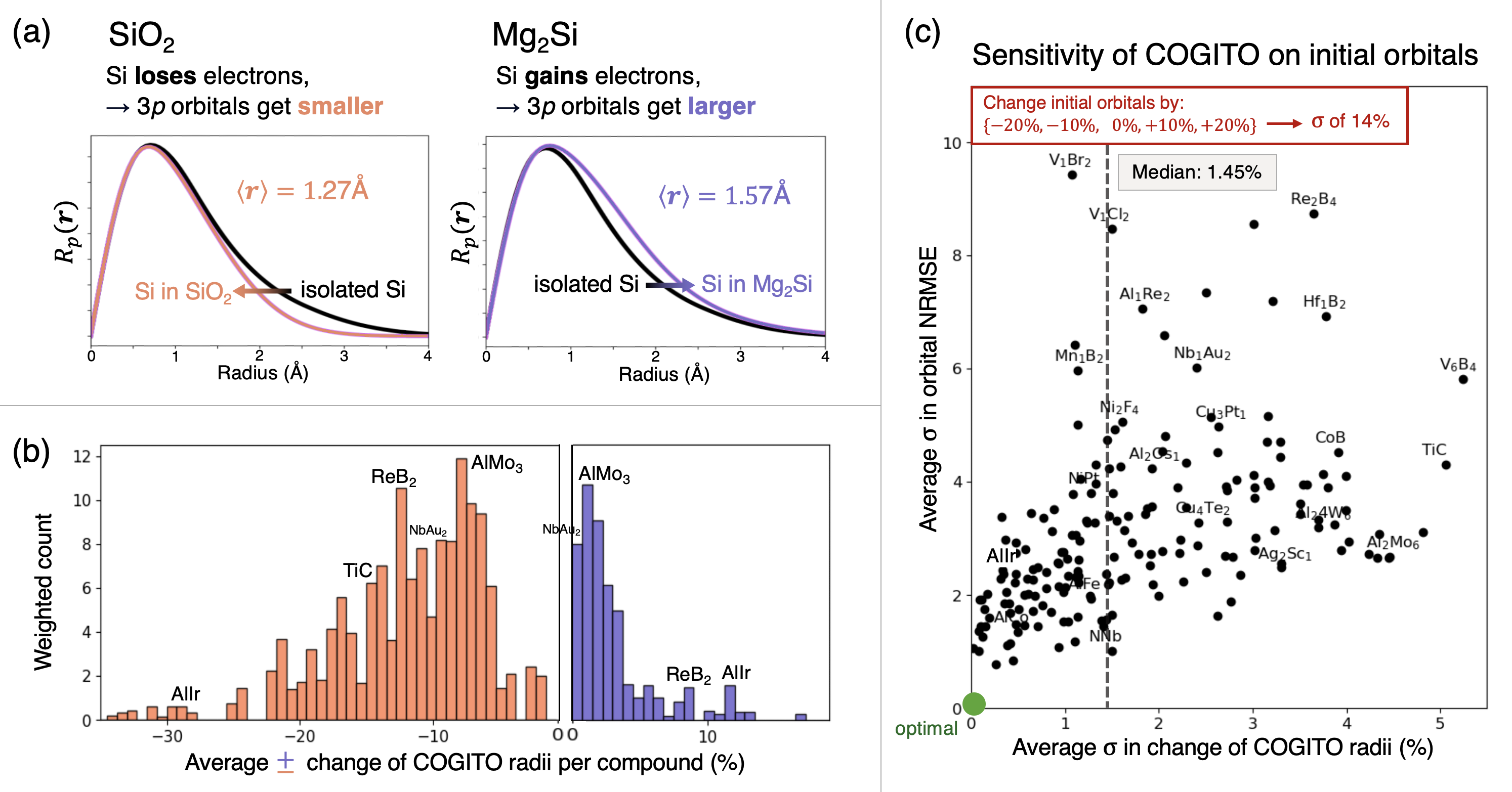}
\caption{Analysis of COGITO radius in SiO\textsubscript{2},
Mg\textsubscript{2}Si, and set of 200 compounds. (a) COGITO radius
changes in SiO\textsubscript{2} and Mg\textsubscript{2}Si consistent with cation/anion behavior of
silicon. (b) Statistical distribution of orbital radius adjustments
across 200 compounds, showing decrease (orange) and increase (blue) in
COGITO radii compared to the PAW pseudo-orbital radii. The histogram is
weighted by the ratio of orbital radii that decrease vs increase for
each compound. A few key compound positions are labeled. For example,
COGITO causes \(\frac{8}{13}\) of the orbitals in AlIr to contract by an
averaged 29\%, while the other \(\frac{5}{13}\) of orbitals expand by an
averaged 13\%. This adds to the plot a bar of height 0.62 at -29\% and
0.38 at 13\%. (c) Sensitivity analysis demonstrating minimal dependence
of COGITO orbital radii on projected atomic basis.}
\label{fig:compare_rad}
\end{figure*}

We demonstrate that COGITO builds high-quality atomic orbitals in four
key manners. (1) A representative example of COGITO radii properly
displays contraction or expansion, depending on the cationic or anionic
character of the ion changes, respectively. (2) Statistical analysis of
the 200 compounds shows substantial and diverse changes in
COGITO-derived orbital radii, indicating the necessity of the COGITO
process compared to a direct projection onto PAW pseudo-orbitals. (3)
Statistical analysis of COGITO radii over a variety of initializations
reveals an 8× reduction in sensitivity to initial orbital size,
demonstrating the robustness of the iterative orbital update approach.
(4) Spurious long-range overlaps are reduced in COGITO by 78\% on
average, indicating the ability of COGITO to preserve a local
description of the orbital chemistry. Altogether, this reflects that the
COGITO basis faithfully captures the electron wavefunction
redistribution from KS-DFT.

First, we illustrate the chemical sensitivity of COGITO by examining how
orbital radii update when ions are cationic vs. anionic---examining
silicon in SiO\textsubscript{2} and Mg\textsubscript{2}Si in the anti-fluorite structure as a
representative example. As a cation in SiO\textsubscript{2}, silicon experiences a
higher effective nuclear charge due to electron loss which reduces the
Si orbital radii. Conversely, silicon behaves as the anion in Mg\textsubscript{2}Si,
yielding a lower effective nuclear charge which increases the Si orbital
radii. COGITO correctly captures these chemical environments, as seen in
\textbf{Fig.~\ref{fig:compare_rad}a}, showing a decreased COGITO radius for Si 3\emph{p} in
SiO\textsubscript{2} and increased COGITO radius for Si 3\emph{p} in
Mg\textsubscript{2}Si. Such changes demonstrate COGITO\textquotesingle s
capacity to capture shifts in atomic orbital properties driven by
different local charge environments and crystal field effects,
establishing a meaningful atomic orbital basis that faithfully
represents the underlying physics of electron redistribution in varying
structural contexts.

Second, we perform a statistical analysis of COGITO-derived changes in
orbital radii across the 200 compounds from Vitale \emph{et al.},
plotted in \textbf{Fig.~\ref{fig:compare_rad}b}. Based on ionic electron transfer causing
expansion or contraction of the cation or anion orbitals, we may expect
an equal number of orbitals to expand as contract, but COGITO reveals
much more is contributing the orbital size. To visualize the effects of
COGITO, we split our data into two sets: COGITO radii that get smaller
(orange) or larger (blue) than the initial PAW pseudo-orbital radii.
Then the orbitals are grouped by compound, and their average radii
change is plotted on the histogram in \textbf{Fig.~\ref{fig:compare_rad}b} with a height
corresponding to the fraction of orbitals that get smaller/larger.
Across all 200 compounds, ¾ of orbitals contract by an averaged 12.4\%
while the remaining ¼ expand by an averaged of 2.8\%. Thus overall,
COGITO reveals a significant preference towards localizing orbitals.

Deviation from the ionic-based expectation of balanced contraction and
expansion is anticipated from covalent bonds and crystal field
repulsion. The formation of covalent bond removes electrons from their
atoms, creating larger effective nuclear charge that more tightly binds
the atomic orbitals. Additionally, interaction of the atomic orbitals
with an exterior Coulomb potential from the surrounding nuclei and
electrons will repeal the electrons, causing atomic orbitals to
contract. COGITO captures all of this intricate physics and chemistry
seamlessly, quickly unveiling dynamics at play in the electronic
structure.

\begin{figure}[hbt!]
\centering
\includegraphics[width=3.4in]{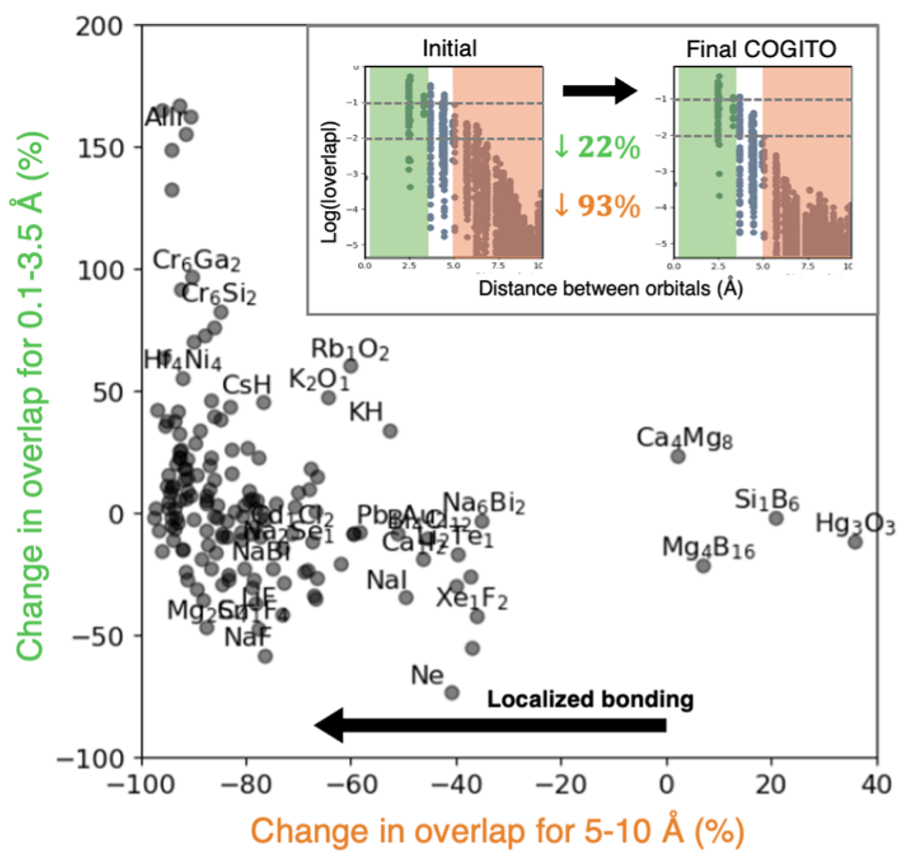}
\caption{The change in long-range (5--10 Å) vs short-range (0.1--3.5 Å)
overlap parameters between the initialized orbitals (PAW pseudo with
exponentially fit tail) and the COGITO basis. The embedded plots show
the log(\textbar overlap\textbar) vs distance between the orbitals (Å).
Bonding metrics become more localized as the long-range overlaps are
reduced. COGITO yields a 78\% average reduction in long-range orbital
overlaps, better decomposing the electron density into short range
interactions.}
\label{fig:compare_decay}
\end{figure}

Third, to test the extent that COGITO is a unique representation of the
DFT-converged electron wavefunctions, we analyze the sensitivity of the
COGITO radii to perturbations in initial orbital radii. This is a
crucial step, as other nonorthogonal orbital constructions that do not
iteratively refine orbital bases, such as QUAMBO/QO or NGWF, display
great sensitivity to their initial conditions. Here, to perturb the
initialization, the initial orbitals (PAW pseudo-orbitals) were
uniformly compressed or expanded by increments of 10\%, ranging from
$-$20\% to +20\%. The sensitivity of COGITO to these perturbations is
quantified by the standard deviation in the percent change of the COGITO
radii from the PAW pseudo-orbital radii. This standard deviation,
plotted on the \emph{x}-axis of \textbf{Fig.~\ref{fig:compare_rad}c}, represents how
consistently COGITO determines the orbital radii irrespective of the
starting conditions. Across all 200 analyzed compounds, the average
standard deviation in the percent change of COGITO radii is 1.45\%. This
deviation is 10\(\times\) smaller than the standard deviation of 14\% in
the initial set of orbitals when initialized with ±20\% variations. The
minimal sensitivity achieved by COGITO underscores its capability to
converge on an orbital basis, substantially mitigating common issues of
gauge freedom and dependence on initial guesses. This robustness is
essential when aiming to deploy COGITO for high-throughput DFT
calculations, as the initialization does not need to be specially
tailored for each chemical system observed.

Additionally, we measure how changes in the COGITO basis from varying
initialization affect the quality of the orbital representation by
calculating the standard deviation of the Normalized Root Mean Square
Error (NRMSE) between the numerical Bloch orbitals and the optimized
COGITO orbitals (y-axis, \textbf{Fig.~\ref{fig:compare_rad}c}). There are some compounds
that exhibit large variations in orbital radii but small variations in
NRMSE (e.g. Al\textsubscript{2}Mo\textsubscript{6}), which suggests an
intrinsic ambiguity in the orbital representation since changes in
orbital size have little effect on descriptions of the KS wavefunctions.
On the other hand, some compounds have relatively large variation in
orbital radii and large variations in NRMSE (e.g. ReB\textsubscript{2}),
such that COGITO can still identify the best orbital representation as
the one which gives the best description of the KS wavefunctions (lowest
NRMSE).

Finally, we examine COGITO's impact on the short-range (0.1-3.5 Å) and
long-range (5-10 Å) overlaps. Long-range overlap terms often arise
spuriously, complicating tight-binding models and reducing
interpretability. Their energy counterpart, long-range hopping terms or
TB parameters, occur frequently in MLWF, PDWF, QO, and LOBSTER as a
result of the long-range oscillating orbital tails shown in the top of
\textbf{Fig.~\ref{fig:distort}}. To reduce both long-range overlap and long-range
hopping, the atomic orbital basis needs to be appropriately localized
while ensuring good projection on the KS wavefunction, exemplified by
low orbital mixing (shown later in \textbf{Fig.~\ref{fig:spill}}).

To analyze how COGITO changes orbital overlap from PAW pseudo-orbitals, we find the percentage change for each overlap term \(S_{\beta\alpha}^{\mathbf{R}}\) and plot the average percent change in the short-range (green) vs. the long-range (orange) region in \textbf{Fig.~\ref{fig:compare_decay}}. The \textbf{R}-dependent overlap matrix is constructed from the Fourier transform of \textbf{k}-dependent overlap matrices, similar to \textbf{Eqn. 37}, and is normalized such that onsite terms are one. 
Overlaps which are $<0.001$ for short-range and $<0.0001$ for long-range are discarded to reduce noise. 
The effect of COGITO on short-range bonds is highly variable, ranging from $–73\%$ to $+165\%$, with the average change at $+11\%$. This demonstrates the adaptability of COGITO. 
Additionally, the magnitude of orbital overlap parameters between 5–10 Å decrease significantly, with half of the compounds reducing by over $85\%$, reflecting increased localization of the atomic basis. 
There are four outlying cases where long-range overlaps increase, Ca\textsubscript{4}Mg\textsubscript{8}, Mg\textsubscript{4}B\textsubscript{16}, SiB\textsubscript{6}, and Hg\textsubscript{3}O\textsubscript{3}. COGITO identifies unusual bonding motifs in these four compounds, which is congruent with their distinctly long-range, multi-center, or low-dimensional bonding. 
Broadly speaking, COGITO’s ability to minimize long-range interactions not only simplifies the computational model but also ensures a clear physical interpretation by accurately representing electron interactions predominantly within short-range distances.

\section{Completing the COGITO Hamiltonian}
\label{sec:ham}

Finally, we build a local Hamiltonian in the COGITO basis by transforming the KS energies and wavefunctions.
Moreover, this effective tight-binding model enables us to build a real-space description of
covalent bonding from the DFT-derived wavefunctions. To start, we project the KS wavefunctions onto the COGITO
basis and expand our projected coefficients from the irreducible
\textbf{k}-point grid to the full Brillouin zone. Then, we optimize the
coefficients and construct the overlap and Hamiltonian matrices in the COGITO basis.

\subsection{Projection of KS wavefunctions on COGITO basis}
\label{sec:ham_proj}

First, we briefly review the PAW and plane-wave formalism to establish
how to project the KS wavefunctions from VASP. Using PAW requires that
we transform our pseudo wavefunction into the all-electron wavefunction
via the transformation operator below. The \textbf{k}-indices are dropped for readability.

\begin{equation}
\mathcal{T}
= 1 + \sum_{i}\Big(\ket{\varphi_{i}}-\ket{\widetilde{\varphi}_{i}}\Big)\bra{p_{i}}.
\tag{23}
\end{equation}

We define our orbital basis to have the same transformation from the
pseudo to all-electron basis as the KS wavefunctions, which leads to
\textbf{Eqn. 24}. 

\begin{align}
\braket{\Phi_{\beta}|\psi_{n}}
&= \braket{\widetilde{\Phi}_{\beta}|\mathcal{T}^{\dagger}\mathcal{T}|\widetilde{\psi}_{n}}\nonumber\\
&= \braket{\widetilde{\Phi}_{\beta}|\Big(1+\sum_{ij}\ket{p_{j}}Q_{ij}\bra{p_{i}}\Big)|\widetilde{\psi}_{n}}.
\tag{24}
\end{align}

Thus, the coefficients in the all-electron basis are:

\begin{equation}
c_{\alpha n} = \sum_{\beta} {S_{\beta\alpha}}^{-1}\,\braket{\Phi_{\beta}|\psi_{n}}.
\tag{25}
\end{equation}

where the orbital overlap is also calculated with the transformation
operators. Crucially here, the pseudo-orbital overlap is not the
identity matrix, as atomic orbitals on different atoms will be
overlapping.

\begin{equation}
S_{\alpha\beta}
= \braket{\Phi_{\alpha}|\Phi_{\beta}}
= \braket{\widetilde{\Phi}_{\alpha}|\widetilde{\Phi}_{\beta}}
+ \sum_{ij}\braket{\widetilde{\Phi}_{\alpha}|p_{j}}\,Q_{ij}\,\braket{p_{i}|\widetilde{\Phi}_{\beta}}.
\tag{26}
\end{equation}

Since VASP uses the plane-wave representation
\(\ket{\mathbf{k} + \mathbf{G}}\), all the overlaps
are computed in Fourier space with the KS pseudo-wavefunctions written
as

\begin{equation}
\ket{\widetilde{\psi}_{n}^{\mathbf{k}}}
= \sum_{\mathbf{G}} c_{\mathbf{G}n}^{\mathbf{k}}\ket{\mathbf{k}+\mathbf{G}}.
\tag{27}
\end{equation}

where \(c_{\mathbf{G}n}^{\mathbf{k}}\) are the plane-wave coefficients
output from VASP. The plane-wave basis projected in real-space is
\(\frac{1}{\sqrt{\Omega}}e^{i\left( \mathbf{k} + \mathbf{G} \right) \cdot \mathbf{r}}\),
where \(\Omega\) is the volume of the primitive unit cell.

A Bloch atomic orbital is written in Fourier space as:

\begin{equation}
\ket{\Phi_{\alpha}^{\mathbf{k}}}
= \frac{4\pi}{\sqrt{\Omega}}\sum_{\mathbf{G}}
e^{- i\mathbf{G}\cdot \bm{\tau}_{\alpha}}\,
\mathcal{F}_{\alpha}(\mathbf{k}+\mathbf{G})\,\ket{\mathbf{k}+\mathbf{G}}.
\tag{28}
\end{equation}

where the phase factor
\(e^{- i\mathbf{G} \cdot \bm{\tau}_{\mathbf{\alpha}}}\) encodes
the orbital center \(\bm{\tau}_{\alpha}\) without numerical error
and \(\mathcal{F}_{\alpha}\left( \mathbf{k} + \mathbf{G} \right)\) is
the analytical Fourier transform of a local atomic orbital as defined by
\textbf{Eqn. 29}.

\begin{align}
\mathcal{F}_{\alpha}(\mathbf{k}+\mathbf{G})
&= \int \phi_{\alpha}(\mathbf{r})\,e^{- i(\mathbf{k}+\mathbf{G})\cdot \mathbf{r}}\,d\mathbf{r}\nonumber \\
&= (-i)^{l}\,Y_{m}^{l}(\mathbf{k}+\mathbf{G})\,K_{\alpha}^{l}(\lvert\mathbf{k}+\mathbf{G}\rvert).
\tag{29}
\end{align}

Here, \(K_{\alpha}^{l}\left( |\mathbf{k}\mathbf{+ G}| \right)\) is an
integral with the radial part of the orbital in real space
\(R_{\alpha}^{l}(r)\) and spherical Bessel functions \(j_{l}\). Using
the gaussian representation of the COGITO basis from \textbf{Eqn. 22}, the
integral can be solved analytically as in the right most side of \textbf{Eqn. 30}
and \textbf{Eqn. 31}.

\begin{align}
K_{\alpha}^{l}\!\left(\lvert\mathbf{k}+\mathbf{G}\rvert\right)
&= \int R_{\alpha}^{l}(r)\,j_{l}\!\left(\lvert\mathbf{k}+\mathbf{G}\rvert r\right)\,r^{2}\,dr\nonumber\\
&= \lvert\mathbf{k}+\mathbf{G}\rvert^{l}\sum_{i}A_{\alpha}^{i}\,e^{-B_{\alpha}^{i}\lvert\mathbf{k}+\mathbf{G}\rvert^{2}},
\tag{30}
\end{align}

\begin{align}
A_{\alpha}^{i} = 2^{-2-l}\sqrt{\pi}\,a_{\alpha}^{i}\,(b_{\alpha}^{i})^{- \frac{3}{2}-l},
\ \ \ \
B_{\alpha}^{i}
= \frac{1}{4b_{\alpha}^{i}}.
\tag{31}
\end{align}

where \(a_{\alpha}^{i}\) and \(b_{\alpha}^{i}\) are previously defined
in \textbf{Eqn. 22} to create \(R_{\alpha}^{l}(r)\).

For a more detail on how we arrive at \textbf{Eqns. 28} and \textbf{29}, we refer readers
to Appendix B of Ref. \cite{agapito_accurate_2016}. Although, our representation differs
slightly from Ref. \cite{agapito_accurate_2016} because we elect to include the phase factor
\(e^{i\mathbf{k} \cdot \mathbf{\tau}_{\alpha}}\) in the creation of
the real-space Bloch atomic orbital, which cancels the phase factor
\(e^{- i\mathbf{k} \cdot \mathbf{\tau}_{\alpha}}\) that arises from
shifting orbital center in the Fourier representation. The real-space
phase factor is then explicitly included in the Fourier transform of
$H$ or $S$ to get local versions in \textbf{Eqn. 37}. We find that this
encoding of the phase factor explicitly in the Bloch orbitals, overlaps,
and Hamiltonians reduces error of the tight binding interpolation.

The PAW projectors in Fourier space are represented the same as the
equations above for the COGITO basis but the radial part
\(K_{\alpha}^{l}(k)\) is found by interpolating the reciprocal radial
part provided by VASP in the POTCAR. Because these all have the same
representation in the orthogonal basis of
\(\ket{\mathbf{k} + \mathbf{G}}\) plane-waves, the
integral of these functions is simply the dot product of their
plane-wave coefficients. Then the overlaps and coefficients are
constructed as defined above.

\subsection{Symmetrize to full BZ from irreducible BZ}
\label{sec:ham_sym}

Once the coefficients of \textbf{k}-points on the irreducible Brillouin
zone are found, they must be expanded to the full Brillouin zone. To
complete this process, we apply the symmetry operations of the crystal
to the reduced \textbf{k}-point and determine if the transformed
\textbf{k}-point is a point on the full Brillouin zone. Once the full
Brillouin zone is reconstructed in terms of reduced \textbf{k}-points and
corresponding symmetry operations, we apply the symmetry operations to
the orbital coefficients of the \textbf{k}-point to get the LCAO
wavefunction for the new \textbf{k}-point.

\subsection{Construct overlap and Hamiltonian matrices}
\label{sec:ham_ham}

Finally, we discuss the best approach for constructing Hamiltonian matrix elements in the COGITO basis. 
While Hamiltonian matrix elements in the COGITO basis may be
computed via direct projection onto the KS wavefunctions,

\begin{equation}
H_{\alpha\beta}^{\mathbf{k}}
\equiv \braket{\Phi_{\alpha}^{\mathbf{k}}|\widehat{H}|\Phi_{\beta}^{\mathbf{k}}}
= \sum_{n} \braket{\Phi_{\alpha}^{\mathbf{k}}|\psi_{n}^{\mathbf{k}}}\,\varepsilon_{n}^{\mathbf{k}}\,\braket{\psi_{n}^{\mathbf{k}}|\Phi_{\beta}^{\mathbf{k}}}.
\tag{32}
\end{equation}

this approach inherits the limitations of an incomplete basis, where
band spillage and orbital mixing leads to reduced accuracy---even within
the valence bands. To circumvent these issues, we instead reconstruct
the Hamiltonian using the coefficient matrices \(c_{in}^{\mathbf{k}}\)
(\textbf{Eqn. 4}) and the COGITO overlap matrices \(S_{\alpha i}^{\mathbf{k}}\),
yielding:

\begin{equation}
H_{\alpha\beta}^{\mathbf{k}}
= \sum_{ijn} S_{\alpha i}^{\mathbf{k}}\,c_{in}^{\mathbf{k}}\,\varepsilon_{n}^{\mathbf{k}}\,
{c_{jn}^{\mathbf{k}}}^{\dagger}\,S_{j\beta}^{\mathbf{k}}.
\tag{33}
\end{equation}

This expression enables the use of coefficient optimization strategies
introduced in \textbf{Sec.~\ref{sec:create_opt}} to exactly reproduce the KS band energies when
diagonalizing \(H_{\alpha\beta}^{\mathbf{k}}\) for bands within the
subset of bands whose overlap matrix is set to be identity. Then, when
the atomic basis describes this identity region with high fidelity (e.g.
\(P_{n}\) \textless{} 2\%), we expect the tight-binding interpolation to
closely match the KS bands.

To define the scope of this optimization, we categorize the KS bands
based on their energies relative to the Fermi level \(E_{F}\). Bands
with \(\varepsilon_{n} \leq E_{F} + 2\ eV\) are deemed the low bands,
which are fully within the identity region and subjected to the full
orthonormalization scheme described in \textbf{Eqns. 17} and \textbf{18}. Bands in the
intermediate range \(E_{F} + 2eV < \varepsilon_{n} \leq E_{F} + 5eV\)
constitute the transition bands (labeled \emph{T}), where a smooth
mixing of the inside-identity region and outside-identity region
orthogonalization is applied. Bands with \(\varepsilon_{n} > E_{F} + 5\)
are the high bands (labeled \emph{H}), which are fully Gram-Schmidt
orthogonalized to the low bands, and partially Gram-Schmidt
orthogonalized to the transition bands. Setting the transition region
between 2 to 5 eV above \(E_{F}\) is arbitrary, but we find these
parameters to be successful in all studied cases.

Specifically, a mixing parameter \(\delta_{T}\) varies continuously from
$\sim$1 to 0 across the transition bands, gradually reducing
the weight of symmetric orthogonalization of \emph{T} and Gram-Schmidt
orthogonalization of the high bands to the transition bands. The energy scale, $E_{scale}$, is selected as 1 eV. 

\begin{align}
\delta_{T}
&= \frac{\tanh\!\left(\frac{(\varepsilon_{T}-E_{F}-2)+1}{E_{scale}}\right)+1}{2},
\tag{34}\\
\breve{c}_{\alpha T}
&= \check{c}_{\alpha T_{2}}\,(1-\delta_{T}) + \check{c}_{\alpha T_{5}}\,\delta_{T},
\tag{35}\\
\breve{c}_{\alpha H}
&= \check{c}_{\alpha H_{2}} - \frac{B_{HT}}{B_{TT}}\,\breve{c}_{\alpha T}\,\delta_{T}.
\tag{36}
\end{align}

where \({\check{c}}_{\alpha T_{2}}\) and
\({\check{c}}_{\alpha T_{5}}\) are the coefficient matrices for the
transition bands obtained after the procedure of \textbf{Eqns. 17} and \textbf{18} is
performed with the identity region extending up to 2 eV and 5 eV,
respectively. Similarly, \({\check{c}}_{\alpha H_{2}}\) represents
the coefficient matrix for the high bands after GS orthogonalization
(\textbf{Eqn. 18}) to bands below 2 eV. All matrices are \textbf{k}-resolved but
the \textbf{k} indices have been suppressed for clarity. Importantly, our
coefficient optimization to improve band interpolation happens at each
\textbf{k}-point independently without any iteration. Compared to the MLWF
coefficient optimization, where each \textbf{k}-point coefficient is
iteratively optimized to have maximum overlap with neighboring
\textbf{k}-points, our approach is less complex with higher efficiency.

The optimized \({\breve{c}}_{\alpha n}^{k}\) are then used to construct
the \textbf{k}-resolved Hamiltonian matrix as above. Finally, the
real-space tight-binding Hamiltonian elements are obtained via Fourier
transform:

\begin{equation}
H_{\alpha\beta}^{\mathbf{R}}
= \sum_{\mathbf{k}} H_{\alpha\beta}^{\mathbf{k}}\,
e^{- i\mathbf{k}\cdot(\mathbf{R}-\boldsymbol{\tau}_{\alpha}+\boldsymbol{\tau}_{\beta})}.
\tag{37}
\end{equation}
The overlap parameters are constructed in the same way. 

We find that including the orbital center \(\boldsymbol{\tau}_{\alpha}\) explicitly in the construction of Bloch orbitals, overlap matrices, and Hamiltonian matrices provides more accurate TB interpolations. 
This is because if not set explicitly (Convention I in Ref. \cite{yusufaly2013tight}), the complex phase from the orbital center becomes implicit in the coefficient matrix (Convention II in Ref. \cite{yusufaly2013tight}), which may then be distorted by the partial orthonormalization procedure. 

Additionally, the phase modulation of the coefficient matrix between Conv. I and Conv. II (Eqn. 3.13 in Ref. \cite{yusufaly2013tight}) is clear for an orthogonal basis but not for a nonorthogonal basis. The coefficient matrix for an orthogonal basis is simply the projection of the orthogonal Bloch orbitals with the KS states, such that excluding the phase in the orthogonal Bloch orbital tracks cleanly to the coefficient matrix. 
But for a nonorthogonal basis the coefficient matrix is the dot product of the inverse overlap matrix with the projection matrix (\textbf{Eqn. 4}). This obfuscates the phase difference from the nonorthogonal Bloch orbital with the variable phase differences due to different orbital centers within the orbital overlap matrix. 
Thus we believe that following Conv. I allows us to better clarify the complex phase that each orbital contributes to the wavefunction, and thus correctly track the orbital bonding, which is controlled in part by complex phase alignment \cite{oliphant_why_2025}.

A final detail crucial in performing the correct bonding analysis of a nonorthogonal
basis is to ensure that the band eigenenergies are not shifted in the
DFT code. Unlike an orthogonal model, the hopping parameters in a nonorthogonal tight-binding model are not
invariant to energy shifting. Shifting within the generalized eigenvalue
problem is expressed as \(H\Psi = (\varepsilon - a\mathbb{I)}S\Psi\),
where the eigenenergies that form the diagonal of \(\varepsilon\) are
shifted by constant \(a\). If \(S = \mathbb{I}\), as it is for an
orthogonal basis, the shifting gets placed entirely on the diagonal of
\(H\), thus only shifting the local orbital energies by \(a\). But for
a general \(S\), the shifting is mixed into onsite \emph{and} offsite
\(H\), incorrectly modifying interatomic hopping terms. Plane-wave
DFT codes commonly shift the average potential energy to zero to make integrating the
potential in reciprocal space possible. To construct the correct offsite terms, the band energies
need to be shifted back to include the \emph{G} = 0 term of the
potential energy.

\section{Results of COGITO projections and band interpolation}
\label{sec:results}

To determine the DFT\(\  \leftrightarrow\) COGITO completeness
(\textbf{Criterion 3}) and tight-binding interpolation quality
(\textbf{Criterion 4}) of COGITO, here we present the charge spillage,
orbital mixing, and band distance errors of the 200 insulators and
metals introduced in \textbf{Sec.~\ref{sec:compare}}. We compare COGITO with our construction
of nonorthogonal tight-binding models from the VASP PAW pseudo-orbitals.
Already our tight-binding model from the PAW orbitals appears better than similar
projection constructed models, seen from comparison of silicon
interpolation in \textbf{Fig.~\ref{fig:band_error}} to Fig. 5 in Ref. \cite{marzari_maximally_2012} and Fig. 1a
in Ref. \cite{sanchez-portal_projection_1995}. This improvement is due to PAW as a better basis set and
the use of a nonorthogonal model.

\begin{figure}[tb!]
\centering
\includegraphics[width=3.4in]{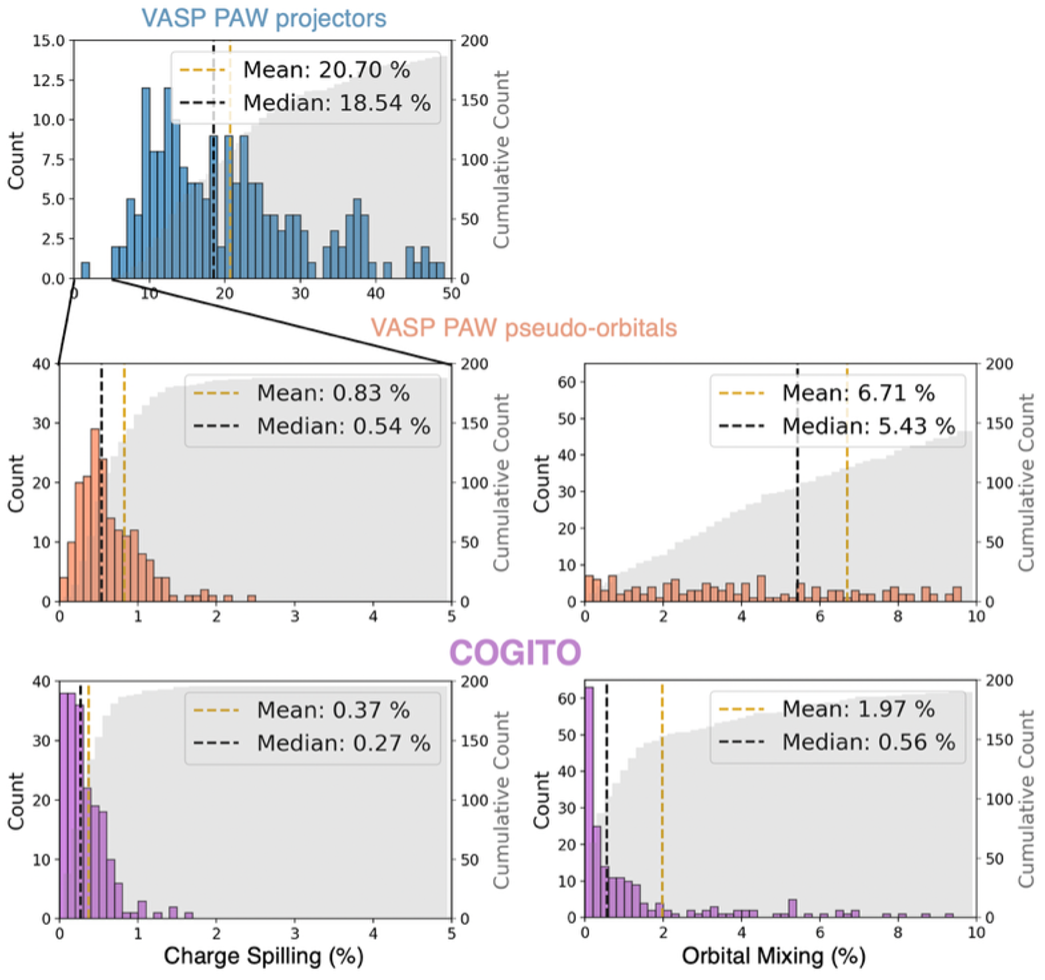}
\caption{Histograms of the 200 compounds for their charge spillage and
orbital mixing. The top row takes the charge spillage from the PROCAR
and LOBSTER output. The bottom panel calculates charge spillage by
projecting the nonorthogonal basis of either the PAW pseudo-orbital or
the COGITO basis.}
\label{fig:spill}
\end{figure}

\begin{figure*}[hbt!]
\centering
\includegraphics[width=7in]{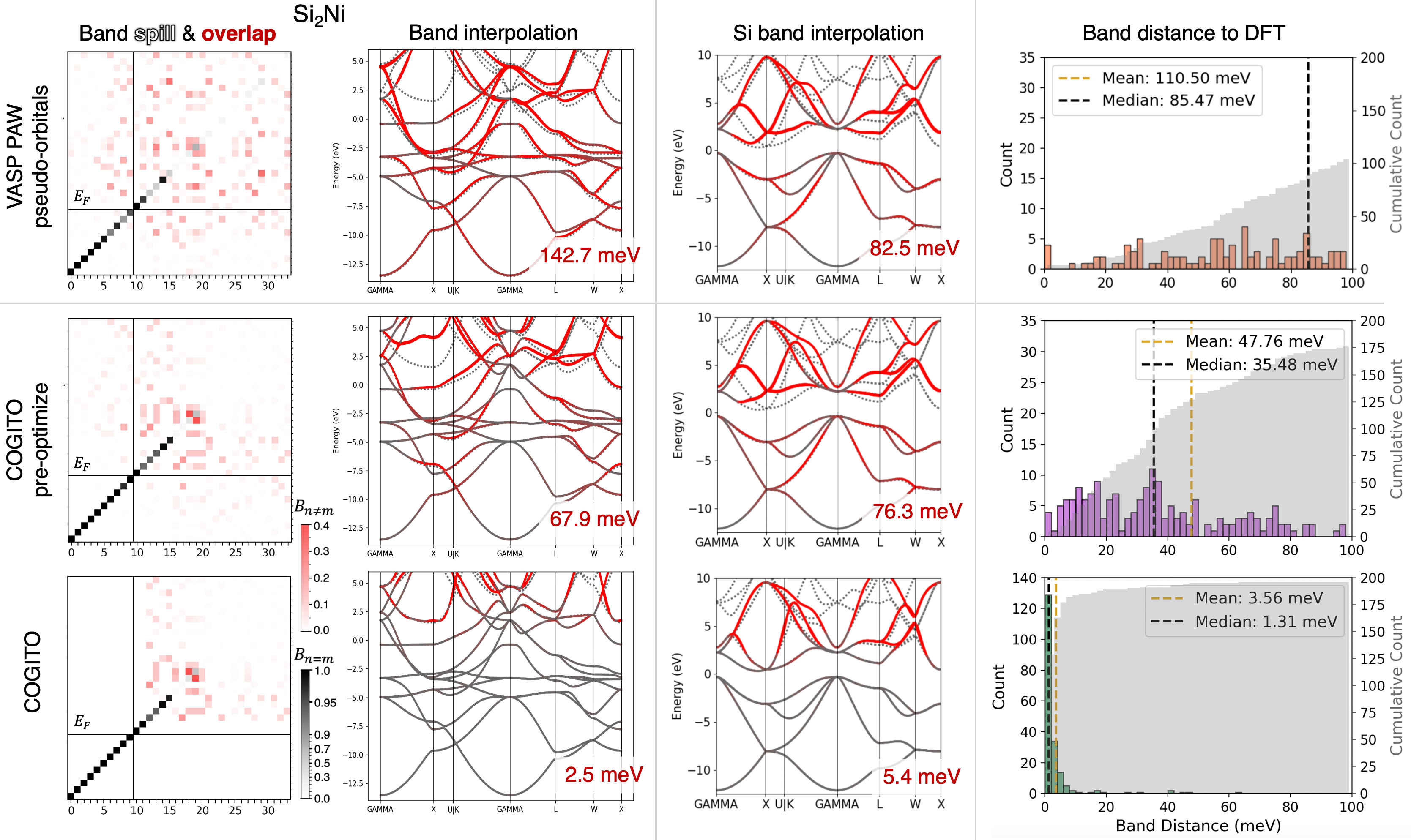}
\caption{Quality metrics for band overlap and band
interpolation error for projection and tight-binding model construction using VASP
PAW pseudo-orbitals, the COGITO basis without the final coefficient
optimization, and the full COGITO construction. The band overlap column
shows that valence band and low-energy conduction bands project better
onto the COGITO basis. The improved projection leads to better band
interpolation of valence band and lowest conduction bands, as
demonstrated by the second two columns. The DFT-calculated bands are
dotted lines and interpolated bands are solid lines that are shaded red as they
deviate from the DFT-calculated bands. The number in the bottom corner
is the band distance (\textbf{Eqn. 38}) for the valence bands. The last column
shows the histogram of band distances for the set of 200 compounds. The
COGITO pre-optimize (middle) reduces the error while the final
coefficient optimization (bottom) hugely reduces the error.}
\label{fig:band_error}
\end{figure*}

We compare the COGITO basis projection with VASP's projection onto PAW
projector functions (performed when LORBIT $\geq$ 10) and our projection onto
VASP PAW pseudo-orbitals. Charge spillage is measured as in \textbf{Eqn.
15} and reflects charge spilled when the occupied KS wavefunctions are
mapped to the projected orbitals. The charge spillage can also be
visualized from the diagonal of the band overlap in \textbf{Fig.~\ref{fig:Bnm_matrix}} as the sum of deviation from solid black for bands below
E\textsubscript{F}. As shown in the top of \textbf{Fig.~\ref{fig:spill}},
the VASP PAW projectors do a poor job of capturing the DFT charge
density, with a median charge spillage of 18.5\% and only one compound
below 5\%. While this is surprisingly high charge-spillage, the PAW
projector's goal of describing how much of the core-region to swap from
pseudo to all-electron may contradict with capturing the overall charge
density. On the other hand, the VASP PAW pseudo-orbitals and COGITO
basis both perform substantially better, with a charge spillage
\textless{} 3\% for all 200 compounds. Either projection is also a
2-5\(\times\) improvement on the charge spillage calculated from
LOBSTER.\cite{naik_quantum-chemical_2023} Compared to the VASP PAW pseudo-orbitals, the COGITO
basis achieves a 2\(\times\) improvement on charge spillage, solidifying
its ability to accurately capture DFT charge density.

Orbital mixing is measured as the maximum off-diagonal component in the
mixing matrix, defined by \textbf{Eqn. 7}. The maximum is taken for each
\textbf{k}-point and averaged to obtain the orbital mixing metric plotted
in \textbf{Fig.~\ref{fig:spill}}. While charge spillage gives a metric for how well
the orbital basis describes the valence band wavefunctions, orbital
mixing reveals how the basis maps onto the KS wavefunctions as a whole.
COGITO achieves an orbital mixing 9× better than the PAW
pseudo-orbitals. The poor PAW orbital mixing is a result of
non-negligible PAW pseudo-orbitals projections on high-energy bands
(\textgreater10 eV above Fermi) not included in the calculated KS bands.
The band overlap matrices in \textbf{Fig.~\ref{fig:band_error}} shows the tendency of PAW
pseudo-orbitals to project on higher energy bands compared to COGITO.
Not only does this result in less accurately capturing low-energy
conduction bands (seen in PAW band interpolation from 0 to 5 eV in
\textbf{Fig.~\ref{fig:band_error}}) but would also require an unfeasible number of bands
to achieve orbital completeness, reducing the quality of the
tight-binding model even for valence bands. COGITO successfully creates
orbitals that project predominantly on the valence bands and low energy
conduction bands, thus obtaining a basis that is complete with less KS
wavefunctions and better reproduces low energy bands.

The tight-binding models are compared to the DFT-calculated band
structure by measuring the band distance error \(\eta_{\nu}\)
(\textbf{Eqn. 38}). The bands to include in \(\eta_{\nu}\) is controlled
by \(f_{n\mathbf{k}}^{\nu}\), which we set as
\(f_{n\mathbf{k}}^{\nu} = 1\) if
\(\varepsilon_{n\mathbf{k}}^{DFT} < E_{F} + \nu\), otherwise as
\(f_{n\mathbf{k}}^{\nu} = 0\). Unless specified otherwise, band distance
is calculated only for the valence bands (\(\nu = 0\)).

\begin{equation}
\eta_{\nu}
= \sqrt{\frac{\sum_{n\mathbf{k}} f_{n\mathbf{k}}^{\nu}\left(\varepsilon_{n\mathbf{k}}^{DFT}-\varepsilon_{n\mathbf{k}}^{TB}\right)^{2}}{\sum_{n\mathbf{k}} f_{n\mathbf{k}}^{\nu}}}.
\tag{38}
\end{equation}

As shown in \textbf{Fig.~\ref{fig:band_error}}, the tight-binding model created from the unmodified
projection of PAW pseudo-orbitals yields an 87.44 meV median band
distance with only 93 compounds below 100 meV. The tight-binding model created from
the unmodified projections (if we were to skip \textbf{Eqns. 34-36}) of COGITO
shows an improvement over the PAW pseudo-orbitals with a 36.76 meV
median band distance and 134 compounds below 100 meV. Once the
coefficient optimization is included, the accuracy of the COGITO band
interpolation is completely transformed, achieving a 1.32 meV median
band distance error with 199 compounds below 100 meV. Overall, the
improved COGITO basis reduces median error by a factor of 2.4 while
COGITO with the coefficient optimization in \textbf{Eqns. 34-36} reduces median
error by a factor of 65. Although it seems like enforcing the
coefficient completeness is more important than the basis optimization
for the band interpolation, the basis optimization is crucial for chemical
and physical interpretation of the resulting tight-binding model. In our next manuscript, we show how the PAW tight-binding
model yields unintuitive bonding results while COGITO succeeds.

\begin{figure}[tb]
\centering
\includegraphics[width=3.2in]{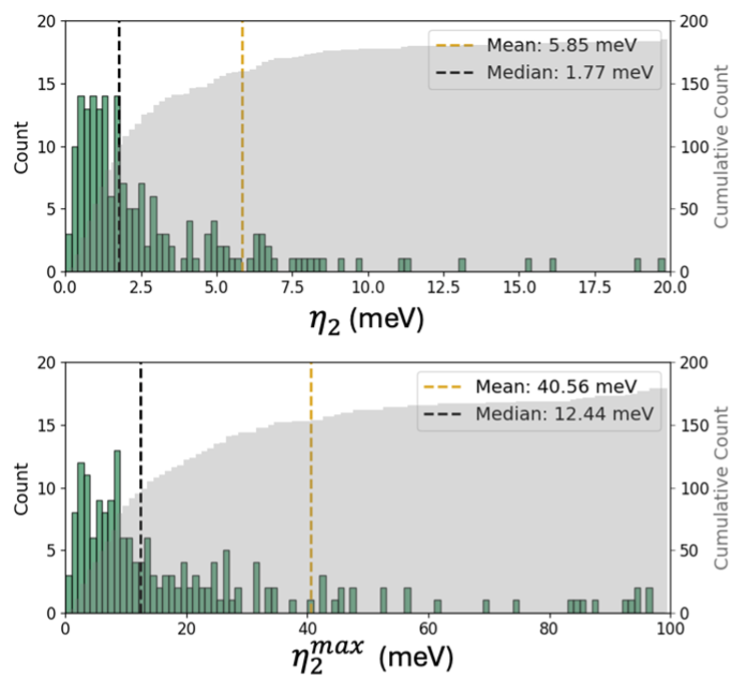}
\caption{The band distance and maximum band error up to
E\textsubscript{F} + 2 eV for the 197 compounds ran with COGITO.}
\label{fig:band_error_pdwf}
\end{figure}

When comparing band distance and maximum band error of COGITO to
Projectability Detangled Wannier Functions (PDWFs)\cite{qiao_projectability_2023}, which is
the newest rendition of MLWFs, COGITO achieves nearly identical quality,
showing its success in accurate electronic structure interpolation. Fig.
5 in Qiao \emph{et. al.} shows histograms of the band distance and
maximum band error for up to E\textsubscript{F} + 2 eV (\(\nu = 2\))
from PDWF and from selected columns of the density matrix
(SCDM)\cite{damle_compressed_2015,damle_scdm-k_2017} for the set of 200 compounds. \textbf{Figure~\ref{fig:band_error_pdwf}} plots
data generated from COGITO in the same format for comparison. They
report that the median value for PDWF is a band distance of 1.60 meV and
a maximum band error of 11.64 meV, whereas SCDM yields 4.80 meV and
33.25 meV, respectively.

\begin{table}[tbh]
\caption{\label{tab:cutoff-data}Data for median band distance and maximum band error over the compound set with various cutoffs. The $\le 100$ or $\le 500$ indicates that  only compounds with band distance below the value are included. The number of compounds below the cutoff is the $N$ column.}
\begin{ruledtabular}
\begin{tabular}{ccccc}
 & \multicolumn{2}{c}{COGITO} & \multicolumn{2}{c}{PDWF~\cite{qiao_projectability_2023}} \\
\cline{2-3}\cline{4-5}
 & $\eta$ (meV) & $N$ & $\eta$ (meV) & $N$ \\
\hline
$\eta_{2}^{\mathrm{median}}$          & 1.77   & 196 & 1.597   & 200 \\
$\eta_{2}^{\mathrm{max,med}}$         & 12.44  & 196 & 11.642  & 200 \\
$\eta_{0}^{\le 100}$                  & 3.56   & 196 & 2.685   & 200 \\
$\eta_{2}^{\le 100}$                  & 4.87   & 195 & 4.231   & 200 \\
$\eta_{4}^{\le 100}$                  & 26.85  & 166 & 22.701  & 179 \\
$\eta_{0}^{\mathrm{max},\le 500}$     & 16.48  & 196 & 20.392  & 200 \\
$\eta_{2}^{\mathrm{max},\le 500}$     & 30.98  & 195 & 32.038  & 198 \\
$\eta_{4}^{\mathrm{max},\le 500}$     & 150.05 & 130 & 132.687 & 152 \\
\end{tabular}
\end{ruledtabular}
\end{table}

Here, we find that COGITO produces medians of 1.77 meV for band distance
and 12.44 meV for maximum band error when including bands up to
E\textsubscript{F} + 2 eV. Overall, COGITO shows a 2.7× decrease in
interpolation error compared to SCDM and 1.1× increase compared to
PDWFs. While these numbers are roughly comparable as they are from the
same set of 200 compounds, there are differences in the workflow that
may introduce mild changes in the error analysis, primarily in that
COGITO uses VASP while Ref {[}36{]} uses Quantum Espresso. This puts
COGITO on the same level of PDWF for band interpolation quality
(Criterion 4) but further affords reliable chemical interpretation from
its adaptable atomic basis (Criterion 1 and 2).
\vspace{-10pt}

\section*{Conclusion}

Here, we showed how projected Wannier orbitals deform from their
original atomic basis, which revealed orbital mixing and an approximate
fixed-overlap constraint as fundamental obstacles of nonorthogonal
projected Wannier methods. To overcome these challenges, we developed
Crystal Orbital Guided Iteration To atomic-Orbitals (COGITO)---an
iterative scheme that co-evolves a strictly atomic basis and its Wannier
representation while enforcing completeness of the target KS
wavefunctions. This establishes a direct and physically grounded
connection between plane-wave DFT and a chemically interpretable atomic
basis.

COGITO's ability to find a chemically-reliable 
atomic basis is grounded in its satisfaction of our four criteria
from the introduction across a test set of 200 compounds. \textbf{(1)}
The minimal COGITO basis is strictly atomic orbitals and decays without
undesirable oscillations or nodes. \textbf{(2)} The COGITO basis adapts
to local atomic environment with limited sensitivity to initialization.
\textbf{(3)} The COGITO basis achieves near completeness\textbf{,} with
a median charge spillage of only 0.27\% and median orbital mixing of
only 0.56\%. \textbf{(4)} The COGITO tight-binding model accurately
reproduces the band structure with a median band distance of 1.3 meV.

In contrast, existing methods for basis construction fall short of these
criteria. Orthogonal Wannier functions (MLWF) delocalize onto
neighboring atoms (\textbf{Fig.~\ref{fig:distort}}), leading to Hamiltonian matrix
elements whose signs and magnitudes no longer align with the underlying
atomic orbital. Nonorthogonal Wannier approaches
(QUAMBO/QAO) are highly sensitive to initialization and often lose
locality due to the fixed overlap constraint (\textbf{Fig.~\ref{fig:distort}}). Fixed
atomic bases, such as used in LOSBTER or VASP projections, neglect
orbital relaxation within the crystal environment, resulting in larger
charge spillage, more orbital mixing, and worse band interpolation.
Moreover, the choice of fixed basis strongly affects projected
quantities. 

COGITO establishes a new foundation for understanding and modeling the
quantum mechanics of atomic orbital interactions in materials. Its minimal and localized yet
complete basis bridges the gap between atomic and plane-wave
representations, offering an ideal framework for many-body extensions
from DFT+U and DMFT to embedding and machine-learned Hamiltonians.
Looking forward, this unified orbital language could enable predictive
models that connect bonding, structure, and function across all classes
of materials---transforming how we interpret and design the electronic
structure of matter.

\bibliography{COGITOrefs_prb}% Produces the bibliography via BibTeX.

\end{document}